\documentclass[conference,compsoc]{IEEEtran}
\usepackage[nocompress]{cite}
\usepackage{amsmath,amssymb}
\usepackage{pifont}
\usepackage[nointegrals]{wasysym}
\usepackage{graphicx}
\usepackage{booktabs}
\usepackage[table]{xcolor}
\usepackage{xspace}
\usepackage{multirow}
\usepackage{placeins}
\usepackage{array}
\usepackage[caption=false,font=footnotesize]{subfig}
\usepackage[hidelinks]{hyperref}
\usepackage{tcolorbox}

\definecolor{ebBlue}{HTML}{DCE3F0}
\definecolor{ebAmber}{HTML}{F4CC79}
\definecolor{ebBeige}{HTML}{ECD7BC}
\definecolor{ebCyan}{HTML}{CCFDFE}
\definecolor{ebCoral}{HTML}{ED8683}
\definecolor{ebPink}{HTML}{F7D3D2}
\definecolor{ebWarm}{HTML}{F2F2EE}
\definecolor{ebDeepBlue}{HTML}{B9C6E0}
\definecolor{ebInk}{HTML}{2E5BE6}

\newcommand{\jev}{Jev-1.13.0\xspace}
\newcommand{\semif}{SemIf-Qwen3.5-4B\xspace}
\newcommand{\laya}{Laya\xspace}
\newcommand{\deepseek}{DeepSeek-V4.1-Flash\xspace}
\newcommand{\glm}{GLM-5.3-Flash\xspace}
\newcommand{\qwenflash}{Qwen3.8-Flash\xspace}
\newcommand{\qwenjson}{Qwen3.5-4B-JSON\xspace}
\newcommand{\reranker}{MiniLM-Reranker\xspace}
\newcommand{\clf}{DistilBERT classifier\xspace}
\newcommand{\corpus}{EdgeIntent~v1\xspace}

\newcommand{\takeaway}[2]{%
  \par\smallskip\noindent\fcolorbox{black}{ebWarm}{\begin{minipage}{0.96\linewidth}\small
  \colorbox{ebAmber}{\textbf{\textit{Takeaway}}}\hspace{0.4em}\textbf{#1.} #2\end{minipage}}\par\smallskip}

\DeclareRobustCommand{\cmark}{{\color{ebInk}\ding{51}}}
\DeclareRobustCommand{\xmark}{{\color{black!35}\ding{55}}}
\DeclareRobustCommand{\pmark}{{\normalfont\LEFTcircle}}

\definecolor{ghdark}{HTML}{24292F}
\definecolor{hfyellow}{HTML}{FFD21E}
\definecolor{hfink}{HTML}{1F2937}
\newcommand{\linkpill}[5]{%
  \href{#5}{\tcbox[on line,colback=#1,colframe=#1,boxrule=0pt,arc=7pt,
    left=6pt,right=7pt,top=2.5pt,bottom=2.5pt,boxsep=0pt,
    fontupper=\sffamily\upshape\small\color{#2}]{#3\hspace{0.45em}#4}}}

\newenvironment{packeditemize}{
    \begin{list}{$\bullet$}{
            \setlength{\labelwidth}{4pt}
            \setlength{\itemsep}{0pt}
            \setlength{\leftmargin}{\labelwidth}
            \addtolength{\leftmargin}{\labelsep}
            \setlength{\parindent}{0pt}
            \setlength{\listparindent}{\parindent}
            \setlength{\parsep}{0pt}
            \setlength{\topsep}{1pt}}}{\end{list}}

\newcommand{\arxivGithubIcon}{\raisebox{-0.774pt}{\includegraphics[width=8.721pt]{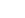}}}
\newcommand{\arxivHuggingFaceIcon}{\raisebox{-0.0pt}{\includegraphics[width=9.0pt]{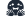}}}
\hypersetup{pdftitle={Replacing Large Language Models with Jev Decision Models for Low-Latency Edge Service Orchestration},
pdfauthor={Delong Li, Xu Wang, Haochen Gong, Rui Lang, Guangsheng Yu}}

\begin{document}
\title{Replacing Large Language Models with Jev Decision Models for Low-Latency Edge Service Orchestration}
\author{\IEEEauthorblockN{Delong Li, Xu Wang, Haochen Gong, Rui Lang, and Guangsheng Yu}
\IEEEauthorblockA{School of Electrical, Mechanical and Biomedical Engineering\\
University of Technology Sydney, Sydney, Australia\\[0.7em]
\linkpill{ghdark}{white}{\arxivGithubIcon}{OniReimu/Edge-Computing-JEV}{https://github.com/OniReimu/Edge-Computing-JEV}\hspace{0.7em}%
\linkpill{hfyellow}{hfink}{\arxivHuggingFaceIcon}{OniReimu/Edge-Computing-JEV}{https://huggingface.co/datasets/OniReimu/Edge-Computing-JEV}}}

\maketitle

\begin{abstract}
Natural-language service requests can require a language-model decision before
execution starts, consuming part of the request's latency budget. We integrate
Jev's decision-oriented application programming interface (API) into edge
service orchestration to reduce this overhead while retaining service
completion. The integration extracts four to eight bounded intent fields and
applies a shared validator, admission policy, and scheduler, accounting for
decision waiting throughout the request timeline. We compare Jev, two
self-hosted decision models, and three hosted large language models (LLMs) on
8{,}280 verified requests and on a live admission path with modeled execution
and a real optical character recognition service. Across 33 test conditions,
Jev reduces median decision latency by 22.7--64.5\% relative to the fastest
LLM. This latency barely moves with input size, contract width, or catalog
size. On
four-field contracts, Jev's API fees per correct decision are 59.7--80.9\% lower
at a cost of a few exact-match points, while wide contracts mark the limit of
the substitution. Receiving the service catalog with each request, Jev names
unseen services as accurately as known ones. On the live admission path, Jev
keeps 0.91--0.95 of requests exact and on time at loads where the LLMs fall
below 0.1. Since caching repeated descriptions gives the interpreters nearly the
same latency, Jev's gain lies in fresh decisions. These results support
decision-model substitution for latency-bound admission on bounded contracts.
\end{abstract}
\begin{IEEEkeywords}
Large language models, decision models, service orchestration, edge computing,
service admission, quality of service.
\end{IEEEkeywords}

\IEEEpeerreviewmaketitle

\section{Introduction}
\label{sec:intro}
\IEEEPARstart{N}{atural-language} interfaces let users describe an edge service
and its execution requirements together. One such request asks the service to
read the text in an image, keep the image at its originating site, and return
the result before a deadline.
Turning that request into an executable job requires both semantic
interpretation and a placement decision. In edge computing, communication,
resource availability, and placement jointly determine whether the service
responds in time~\cite{shi2016edge,mao2017survey,santos2021lowlatency}.
An interpreter on the admission path consumes part of that same response
budget, delaying execution and leaving less time to complete the service.

\begin{figure}[!t]
\centering
\includegraphics[width=\linewidth]{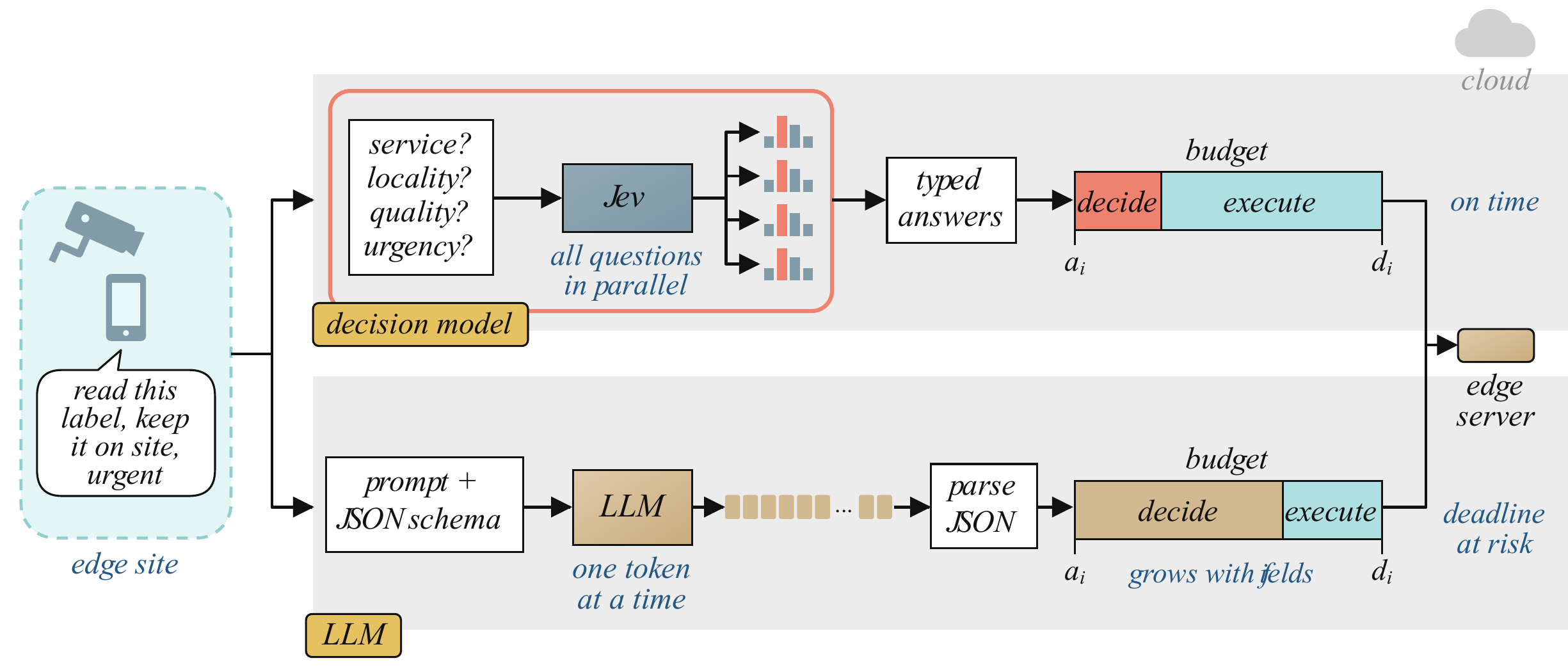}
\caption{Two ways to interpret the same edge request. A camera or phone at an
edge site asks for a service that must run on the site's edge server. A
decision model such as \jev scores every contract question in parallel against
the request text and returns a typed answer with a probability distribution over
the declared options. Because an LLM decodes a JSON object one token at a time,
its decision time grows with the number of fields. The bars sketch how much of the
budget between arrival $a_i$ and deadline $d_i$ each interpreter leaves for
execution.}
\label{fig:overview}
\end{figure}

For a bounded service catalog, interpretation often requires only a small
set of decisions. These decisions include service type, placement permission,
quality tier, and urgency. A large language model (LLM) can express these decisions through
a short structured response, but the downstream scheduler needs the selected
values rather than a generated explanation. A decision-oriented model could
reduce this overhead even relative to a generative model configured for
concise output (Fig.~\ref{fig:overview}). At the service level, the goal is to reduce this decision
overhead while preserving correct, on-time completion under the request's
execution requirements.

Intent-based networking already separates desired outcomes from their
implementation~\cite{clemm2022intentbased}. Language-driven network systems
and service prototypes connect interpretation to configuration and
orchestration~\cite{jacobs2021hey,wang2024netconfeval,angi2025llnet,miyaoka2025chatdriven,nisiotis2026prompt}.
Small models and discriminative extraction offer further ways to specialize
bounded language tasks~\cite{belcak2025small,zaratiana2025gliner2}.
We examine Jev, a decision-oriented service accessed through an application
programming interface (API)~\cite{friedman2026jev}, as the interpreter in an
edge-service admission path. We examine two self-hosted decision models,
\semif~\cite{semif2026openjev} and \laya~\cite{laya2026typed}, in the same role.
Queueing, communication, and execution all
contribute to response time. Their interaction with interpretation determines
how much a faster decision benefits the service.

We investigate five linked questions. The first four isolate the interpreter
and ask how decision latency, semantic correctness, and billed API cost change
as the input grows (RQ1) and as its wording degrades (RQ2). The same three
quantities are measured as the contract gains fields (RQ3) and as the service
catalog grows and changes (RQ4). The fifth
asks whether the substitution lowers full response time while retaining
correct service completion, under what workload conditions faster
interpretation is most useful, and how it interacts with caching for
repeated requests (RQ5).

We address these questions in two layers. The interpretation layer uses
\corpus, a benchmark of 8{,}280 English requests whose labels are fixed first
and whose text is kept only when a blind verifier recovers every label. On it, three decision models face three hosted LLMs
(\deepseek, \glm, and \qwenflash) in every condition. The service layer combines
live interpretation with modeled execution, followed by a real three-node
optical character recognition (OCR) service. The real service transfers
images, executes recognition, and checks the returned text against reference
annotations. A fixed rule parser, a generative model that shares \semif's
weights, a sentence-embedding reranker, and a trained service classifier
extend the comparison beyond the six interpreters.

The contributions are as follows.
\begin{packeditemize}
\item We integrate decision models into an edge-service admission system
through a typed intent contract of four to eight fields with a shared validator
and scheduler. The system places interpretation, queueing, communication, and
execution on one timeline and judges each interpreter by service deadlines and
completion.
\item Three decision models and three hosted LLMs interpret the same contracts
in 33 test conditions. We measure their decision latency, API fees per correct
decision, and accuracy. The conditions vary the input size, the bundling of
requests, and the width of the contract.
\item A catalog test gives each interpreter the service catalog with every
request and checks whether it names services unseen during development as
accurately as known ones, with a frozen service classifier as the reference.
The same test finds the option limit of each decision model.
\item One cache policy, applied to every interpreter, separates the latency of
fresh interpretation from that of repeated requests. The comparison locates
where fast decision models reduce service latency once caching is in place.
\end{packeditemize}

\section{Related Work}
\label{sec:related}
Table~\ref{tab:related} lists the evaluation dimensions that the closest
related studies report, with columns chosen for an interpreter in an edge
admission path. Seven of the fourteen network-intent and orchestration studies
report interpretation latency as a result, and two report its tail. LLNet alone
targets energy, and none of the fourteen varies the offered load or evaluates
reuse of model outputs. Across all twenty-two studies, GLiNER2 alone evaluates
options unseen in training and InferLine alone sweeps the request rate. No
study reports more than six of the eleven dimensions as measured results. This
study measures all eleven for the same interpreters.

\begin{table*}[!t]
\centering
\caption{Evaluation coverage of the closest related work. \cmark{} measured and reported as a result.
\pmark{} mentioned, enforced by design, or reported only for the whole pipeline. \xmark{} not addressed.
Unseen catalog: options absent from training and passed at request time. Safety violations: measured placement,
locality, or policy violations in the decisions. Deadline completion: whether the downstream service completes
correctly or on time. Cov.\ counts \cmark{} cells. $^\dagger$Assessed from the abstract.}
\label{tab:related}
\scriptsize
\setlength{\tabcolsep}{3pt}
\setlength{\aboverulesep}{0pt}\setlength{\belowrulesep}{0pt}\setlength{\extrarowheight}{0pt}
\renewcommand{\arraystretch}{1.05}
\def\ebtab{%
\begin{tabular}{llcccccccccccc}
\toprule
\rowcolor{ebBlue}Work & Paradigm & \ebh{NL$\to$}{typed} & \ebh{Decision}{latency} & \ebh{Tail}{latency} & \ebh{Monetary}{cost} & \ebh{Energy}{or power} & \ebh{Unseen}{catalog} & \ebh{Safety}{violations} & \ebh{Deadline}{completion} & \ebh{Offered}{load} & \ebh{System}{testbed} & \ebh{Output}{reuse} & \ebh{Cov.}{(\cmark)} \\
\midrule
\rowcolor{ebAmber!45}\multicolumn{14}{l}{\textbf{\textit{Network intent interpretation}}} \\
Lumi~\cite{jacobs2021hey} & NER + intent grammar & \cmark & \pmark & \xmark & \xmark & \xmark & \xmark & \xmark & \xmark & \xmark & \cmark & \xmark & 2 \\
Manias et al.~\cite{manias2024intentbased} & Prompted LLM & \cmark & \xmark & \xmark & \xmark & \xmark & \xmark & \xmark & \xmark & \xmark & \xmark & \xmark & 1 \\
NetConfEval~\cite{wang2024netconfeval} & LLM benchmark & \cmark & \pmark & \xmark & \cmark & \xmark & \pmark & \pmark & \pmark & \xmark & \cmark & \pmark & 3 \\
LLNet$^\dagger$~\cite{angi2025llnet} & Fine-tuned SLM & \pmark & \xmark & \xmark & \pmark & \pmark & \xmark & \xmark & \xmark & \xmark & \xmark & \xmark & 0 \\
Lira et al.~\cite{lira2025network} & Fine-tuned SLM & \pmark & \cmark & \pmark & \pmark & \xmark & \xmark & \pmark & \xmark & \xmark & \xmark & \pmark & 1 \\
\rowcolor{ebAmber!45}\multicolumn{14}{l}{\textbf{\textit{Intent-driven service orchestration}}} \\
Brodimas et al.~\cite{brodimas2025intentbased} & Agentic LLM & \pmark & \pmark & \xmark & \cmark & \xmark & \xmark & \pmark & \cmark & \xmark & \cmark & \xmark & 3 \\
JAUNT~\cite{li2025jaunt} & Agentic LLM router & \cmark & \pmark & \xmark & \xmark & \xmark & \pmark & \xmark & \cmark & \xmark & \pmark & \xmark & 2 \\
Miyaoka et al.~\cite{miyaoka2025chatdriven} & Classifier/LLM + ILP & \cmark & \cmark & \xmark & \pmark & \xmark & \xmark & \pmark & \pmark & \pmark & \pmark & \xmark & 2 \\
Intent Engine~\cite{islam2026intent} & Grounded LLM & \cmark & \cmark & \cmark & \pmark & \xmark & \pmark & \cmark & \pmark & \xmark & \pmark & \xmark & 4 \\
Martins et al.~\cite{martins2026intentdriven} & Agentic LLM & \cmark & \cmark & \xmark & \pmark & \xmark & \pmark & \cmark & \xmark & \xmark & \xmark & \xmark & 3 \\
DMO-GPT$^\dagger$~\cite{mekrache2026dmogpt} & Agentic LLM & \pmark & \xmark & \xmark & \xmark & \xmark & \pmark & \xmark & \xmark & \xmark & \cmark & \xmark & 1 \\
Nisiotis et al.~\cite{nisiotis2026prompt} & Fine-tuned SLM router & \cmark & \cmark & \cmark & \pmark & \xmark & \pmark & \pmark & \pmark & \xmark & \cmark & \xmark & 4 \\
Parra-Ullauri et al.~\cite{parra-ullauri2026rolebased} & Agentic LLM & \pmark & \cmark & \xmark & \pmark & \xmark & \pmark & \pmark & \cmark & \xmark & \cmark & \xmark & 3 \\
Seo et al.~\cite{seo2026integrated} & LLM + RL agent & \cmark & \cmark & \xmark & \pmark & \xmark & \xmark & \cmark & \pmark & \xmark & \pmark & \xmark & 3 \\
\rowcolor{ebAmber!45}\multicolumn{14}{l}{\textbf{\textit{Bounded decisions and structured output}}} \\
NetLLM~\cite{wu2024netllm} & Adapted LLM + task head & \xmark & \cmark & \pmark & \pmark & \xmark & \xmark & \pmark & \cmark & \pmark & \cmark & \xmark & 3 \\
GLiNER2~\cite{zaratiana2025gliner2} & Schema-driven encoder & \cmark & \cmark & \xmark & \pmark & \xmark & \cmark & \xmark & \xmark & \xmark & \xmark & \xmark & 3 \\
JSONSchemaBench~\cite{geng2025jsonschemabench} & Constrained decoding & \pmark & \cmark & \xmark & \xmark & \xmark & \pmark & \pmark & \xmark & \xmark & \xmark & \xmark & 1 \\
\rowcolor{ebAmber!45}\multicolumn{14}{l}{\textbf{\textit{Model routing, caching, and serving}}} \\
Clipper~\cite{crankshaw2016clipper} & Serving system & \xmark & \cmark & \cmark & \pmark & \xmark & \xmark & \xmark & \cmark & \pmark & \cmark & \cmark & 5 \\
InferLine~\cite{crankshaw2020inferline} & Pipeline serving & \xmark & \cmark & \cmark & \cmark & \xmark & \xmark & \xmark & \cmark & \cmark & \cmark & \xmark & 6 \\
GPTCache~\cite{bang2023gptcache} & Semantic cache & \xmark & \cmark & \xmark & \pmark & \xmark & \xmark & \xmark & \xmark & \xmark & \xmark & \cmark & 2 \\
FrugalGPT~\cite{chen2023frugalgptb} & LLM cascade & \xmark & \pmark & \xmark & \cmark & \pmark & \xmark & \xmark & \xmark & \xmark & \xmark & \pmark & 1 \\
RouteLLM~\cite{ong2024routellma} & LLM router & \xmark & \pmark & \xmark & \cmark & \xmark & \pmark & \xmark & \xmark & \pmark & \xmark & \xmark & 1 \\
\midrule
\rowcolor{ebCoral!10}\textbf{This work} & Decision models vs.\ LLMs & \cmark & \cmark & \cmark & \cmark & \cmark & \cmark & \cmark & \cmark & \cmark & \cmark & \cmark & \textbf{11} \\
\bottomrule
\end{tabular}}
\def\ebh#1#2{\shortstack{#1\strut\\{}#2\strut}}\sbox0{\ebtab}
\setlength{\tabcolsep}{\dimexpr\tabcolsep+(\linewidth-\wd0-1pt)/28\relax}
\ifdim\tabcolsep<1.5pt\setlength{\tabcolsep}{1.5pt}\fi
\typeout{EBTAB tab_related natural=\the\wd0\space line=\the\linewidth\space sep=\the\tabcolsep}
\ebtab
\end{table*}

\subsection{Intent Interpretation and Service Orchestration}
Intent-based networking separates desired outcomes from the mechanisms used to
realize them~\cite{clemm2022intentbased}. Lumi translates natural-language
network requirements through a learned interface and an intermediate
representation~\cite{jacobs2021hey}. NetConfEval evaluates language models on
network-configuration tasks spanning policy translation, API calls, and
configuration generation~\cite{wang2024netconfeval}. Work on fifth-generation (5G) mobile-network intent
extraction likewise studies the interpretation stage~\cite{manias2024intentbased}.
LLNet and fine-tuned small-model network configuration bring more specialized
models into this setting~\cite{angi2025llnet,lira2025network}. These studies
establish natural-language interpretation as a systems component whose
accuracy and execution overhead both matter.

Several systems extend interpretation into service orchestration.
Chat-Driven Optimal Management of Virtual Network Services combines an
interpreter with optimization~\cite{miyaoka2025chatdriven}. From Prompt to
Service evaluates intent routing and real conversational service paths,
including a fine-tuned small router~\cite{nisiotis2026prompt}. Intent Engine
uses intent grounding and validation within a network
orchestration architecture~\cite{islam2026intent}. Agentic infrastructure
orchestration, zero-touch network pipelines, and DMO-GPT address broader
automation workflows~\cite{brodimas2025intentbased,seo2026integrated,mekrache2026dmogpt}.
JAUNT studies intent-aware network tool routing~\cite{li2025jaunt}, while
grounded and role-based sixth-generation (6G) systems examine additional forms of orchestration
and agent organization~\cite{martins2026intentdriven,parra-ullauri2026rolebased}.

Building on this work on service routing, we compare three decision models
with three generative interpreters while holding the intent contract and
downstream policy fixed.
We measure completion against the returned service output, accounting for
decision waiting, case-sensitive OCR correctness, and caching under the same
policy for each backend. The decision unit is a user-issued service job,
whose admission path can include hundreds of milliseconds of interpretation.

\subsection{Bounded Decisions and Small Models}
Efficient prediction need not require free-form generation. Sentence-BERT
and SetFit support lightweight semantic representations and
classification~\cite{reimers2019sentencebert,tunstall2022efficient}.
GLiNER and GLiNER2 provide discriminative extraction approaches, with GLiNER2
also addressing structured extraction and classification~\cite{zaratiana2024gliner,zaratiana2025gliner2}.
The argument for small models in agentic systems further motivates assigning
bounded tasks to specialized components~\cite{belcak2025small}.
For networking decisions, NetLLM adapts language-model representations to
networking tasks through task-specific components~\cite{wu2024netllm}.

Because a trained classifier fixes its label set at training time, a service added
to the catalog needs labelled examples and a new training run. Decision models
that receive the catalog with each request avoid that step. RQ4 measures the
difference against a DistilBERT service classifier~\cite{sanh2019distilbert}
and a MiniLM sentence-embedding reranker~\cite{wang2020minilm,reimers2019sentencebert}.

Alongside the hosted comparison, we include a fixed rule parser and an open
generative model, Qwen3.5-4B~\cite{qwen2026qwen35}, which shares its weights
with \semif. Rules provide a low-overhead interpretation policy, while Qwen provides
a self-hosted generative alternative under the same service contract. The
two self-hosted arms therefore differ only in how the answer is read out.

\subsection{Serving, Structured Output, and Cost}
Serving systems such as Clipper and InferLine address inference deployment
under latency and resource constraints~\cite{crankshaw2016clipper,crankshaw2020inferline}.
PagedAttention and SGLang improve important aspects of language-model
serving~\cite{kwon2023efficient,zheng2023sglang}. Consequently, a measured
self-hosted model latency is a property of its serving configuration as well
as its weights.

Structured-output validity also needs to be distinguished from task
correctness. JSONSchemaBench evaluates efficiency, constraint coverage, and
quality in structured generation~\cite{geng2025jsonschemabench}. A valid JavaScript Object Notation (JSON)
object can still request the wrong service, just as a correctly interpreted
request can yield incorrect OCR text. We retain these separate endpoints.
Cost-aware prediction and routing systems such as FrugalML, FrugalGPT, and
RouteLLM consider different ways to allocate model calls under quality and
cost objectives~\cite{chen2020frugalml,chen2023frugalgptb,ong2024routellma}.
We compare fixed interpretation backends under a shared admission policy.

Caching is another direct alternative to repeatedly paying inference
latency. GPTCache studies reuse through a semantic cache~\cite{bang2023gptcache}.
Our cache reuses valid parsing results for identical normalized intent text.
Applying the same policy to every backend tests how decision reuse changes
the value of faster interpretation.

\section{Service Admission Architecture}
\label{sec:system}
\begin{figure*}[!t]
\centering
\includegraphics[width=\textwidth]{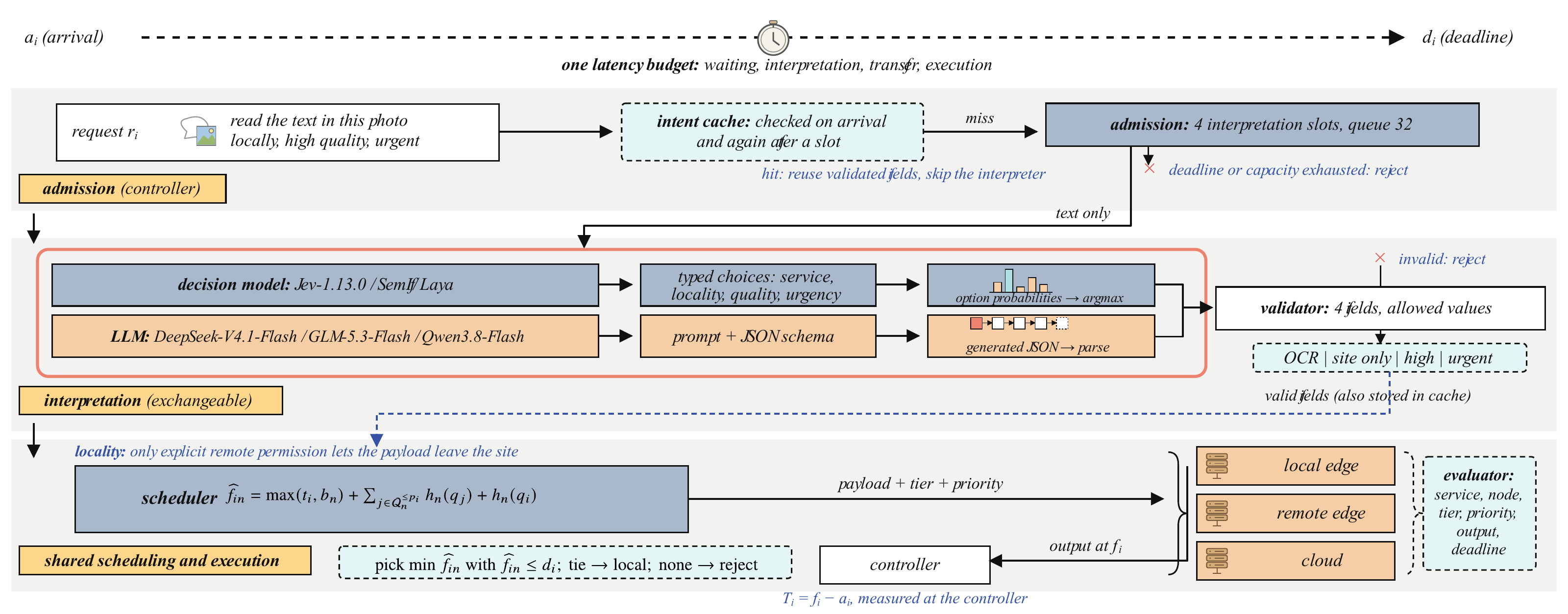}
\caption{Admission path in detail. The controller checks the intent cache on
arrival and again after an interpretation slot frees. The interpreter receives
only the request text and returns typed choices or a JSON object, which the
shared validator checks. The scheduler places each valid request with the
predicted finish rule of Eq.~\eqref{eq:finish}. The evaluator compares service,
node, tier, priority, output, and finish time with the reference requirements.}
\label{fig:architecture}
\end{figure*}
\subsection{Request and Interpretation Contract}
Each request $i$ contains a natural-language description, an arrival time
$a_i$, an absolute deadline $d_i$, an origin, and payload metadata. Times
share the originating controller's clock. Only the
description is sent to the interpreter. The scheduler receives the remaining
metadata through the same interface for every backend. Reference intent
labels and expected service outputs are available exclusively to the
evaluator. Table~\ref{tab:schema} defines the shared interpretation contract.
Its four core fields describe the requested service and execution
requirements. With a catalog of $K$ services, the service field has $K+1$
values, and the core contract hence admits $27(K+1)$ tuples. Two further tiers add
retention and energy ($F=6$), then redundancy and latency class ($F=8$). Each
added field maps to a scheduler action. Missing requirements remain
unspecified rather than being inferred from the evaluator's labels.

\begin{table}[t]
\caption{Shared intent contract. OCR denotes optical character recognition.
$F$ is the number of fields in the contract.}
\label{tab:schema}
\centering
\footnotesize
\begin{tabular}{@{}llp{0.62\linewidth}@{}}
\toprule
Field & $F$ & Allowed values \\
\midrule
Service & 4 & Catalog of $K$ services (count, detection, OCR, \ldots), unsupported \\
Locality & 4 & Site only, remote allowed, unspecified \\
Quality floor & 4 & Standard, high, unspecified \\
Urgency & 4 & Normal, urgent, unspecified \\
Retention & 6 & Discard after use, retain allowed, unspecified \\
Energy & 6 & Eco, performance, unspecified \\
Redundancy & 8 & Single, replicated, unspecified \\
Latency class & 8 & Real time, interactive, batch, unspecified \\
\bottomrule
\end{tabular}
\end{table}

The decision models select among the allowed values, while the generative
backends return the same fields in a concise structured response. The rule-based interpreter
maps the same input text to this contract. The required meaning of each
output is fixed across interfaces. For example, a request to read an image
locally with high quality and urgent handling maps to OCR, site
only, high, and urgent. Selecting the execution node remains the scheduler's
responsibility.

Each contract field becomes one choice question whose criteria text is shared
verbatim with the LLM prompt and schema descriptions. \jev answers all
questions of a request in one hosted call. \semif reads the same questions
out of Qwen3.5-4B weights, and \laya is a 421M-parameter typed-decision
model. Both models run on the edge node. Because neither self-hosted model
generates text, neither carries a per-token decoding cost. The catalog travels with the
request as the options of the service question. A new service therefore
reaches the interpreter without retraining, up to the number of options each
model accepts.

\subsection{Validation, Caching, and Admission}
The validator checks that all required fields are present and that each value
belongs to its allowed set. Invalid interpretations do not proceed to
scheduling. Requests also leave the admission path when their deadline or
the available admission capacity is exhausted. This common policy controls
which responses can be executed. Semantic correctness must be evaluated
separately against the request's intended meaning.

When caching is enabled, each backend reuses its own validated interpretation
for identical normalized request text under the same extraction policy.
A response becomes reusable only after it has arrived. The cache contains
neither reference labels nor service outputs, and it can retain a
well-formed but semantically incorrect interpretation. Because repeated descriptions
may accompany different images, reusing a decision does not eliminate
the corresponding service execution. All backends use the same cache policy.

Admission and execution share one elapsed-time axis. Waiting for an
interpretation slot and receiving a remote response both consume the budget
between $a_i$ and $d_i$. A call that outlives its request's deadline continues
to occupy its slot until the call terminates, while new requests continue
to arrive. Interpretation latency thus affects both service feasibility
and queue occupancy.

\subsection{Shared Scheduling and Execution}
After interpretation, the scheduler reads the current worker state. In the
real service of RQ5 the workers are a local node, a second edge node, and a
cloud node, each with its own link delay and bandwidth. Only an
explicit remote permission allows a payload to leave its originating site.
Unspecified locality defaults to local execution. A high quality requirement
selects the high tier, and other requests use the standard tier. Urgent
requests take priority over ordinary requests without interrupting active
work. These shared defaults can make distinct field values operationally
equivalent, as with normal and unspecified urgency.

For the real service, let $t_i$ denote the time at which request $i$ is
scheduled, after its interpretation has completed. Its selected tier is
$q_i$, and its priority is $p_i$, with smaller values indicating higher
priority. Let $\mathcal{N}_i$ contain the nodes that provide the interpreted
service, satisfy its placement restriction, and have queue capacity.
For a node $n\in\mathcal{N}_i$, let $b_n$ be the estimated finish time of
active work, or zero if idle. Let $h_n(q)$ be the calibrated duration of
serving tier $q$, including transfer and response overhead. Finally,
$\mathcal{Q}_n^{\leq p_i}$ is the set of pending jobs with equal or higher
priority than request $i$, and $q_j$ is pending job $j$'s tier. All node
quantities refer to the state observed at $t_i$. The predicted finish time
$\widehat f_{in}$ is
\begin{equation}
\widehat f_{in}=\max(t_i,b_n)
  +\sum_{j\in\mathcal{Q}_n^{\leq p_i}}h_n(q_j)+h_n(q_i).
\label{eq:finish}
\end{equation}
The scheduler selects the node with the smallest predicted finish time among
those satisfying $\widehat f_{in}\leq d_i$, breaking ties in favor of the
local node. It rejects the request if no such node exists. Equation~\ref{eq:finish}
is a shared admission estimate. Execution-time variation and subsequent
higher-priority arrivals can invalidate its prediction. Correct completion
therefore depends on the observed result and finish time, not merely on
passing the admission test.

An accepted request transfers its payload to the selected worker and returns
the service output to the originating controller. Payloads are transferred
only after node selection. The evaluator checks the actual service, node,
tier, priority, output, and deadline against the reference requirements.
All elapsed times are measured at the originating controller. Locality
constraints govern service payloads, but the textual description still reaches
the chosen interpretation API. The scheduler and these checks are common
to all backends, allowing the comparison to focus on interpreter substitution.

\section{Evaluation Methodology}
\label{sec:design}
\subsection{Comparison and Evidence Design}
The principal comparison is between decision models and LLMs configured for
concise structured output without a generated reasoning trace. The decision
models are \jev (hosted), \semif, and \laya (self-hosted). The LLMs are
\deepseek, \glm, and \qwenflash, which are all hosted and use strict JSON
schemas and temperature zero. All six appear in every condition and use the same intent
contract, validator, admission limits, execution policy, and caching
mechanism. References are evaluated alongside the six without entering their
comparison. \qwenjson generates the JSON object from the same Qwen3.5-4B weights as
\semif, which isolates readout from generation at fixed weights and deployment.
A fixed rule parser serves RQ1--RQ3. RQ4 adds a DistilBERT service classifier,
frozen (DistilBERT-Clf-Frozen) or retrained on the changed catalog
(DistilBERT-Clf-Retrained), and an all-MiniLM-L6-v2 sentence-embedding reranker
(\reranker).
Model identifiers, software, hardware, workload parameters, and execution
procedures are given in Section~\ref{sec:settings}.

RQ1--RQ4 form the interpretation layer. Because each sends identical request
texts to every interpreter and scores the returned fields, only the interpreter
changes between arms. RQ5 forms the service layer in two parts, as in our
earlier exploratory study. Part~A measures live decisions and combines their
actual timing with modeled service execution. It evaluates the interpretation
tradeoff and how decision waiting consumes execution slack. Part~B executes a
real three-node OCR service and checks the returned text. It tests whether
lower decision latency remains useful after actual communication and
processing. The measured hosted latency includes provider execution, routing,
and network transport throughout.

\subsection{The \corpus Benchmark}
\label{sec:corpus}
\corpus fixes each request's label tuple first, with balanced marginals per
field, and has a generator model write the text second. A blind verifier then
labels the text in a separate session without seeing the tuple, and a case is
kept only when its labels match the tuple on every field. Lint checks reject
texts that leak field names or enumeration values. The generators are
Gemini-3.8-Flash and Claude-Opus-5.5 and the verifier is Claude-Opus-5.5. Both
models lie outside the evaluated roster. The texts span six wording families, from operator
tickets to voice-assistant utterances. Each of the 23 generated conditions has
300 test and 60 development cases, 8{,}280 in total. Ten further conditions are
derived programmatically, for example by padding or noise. Development cases
serve only prompt checks, classifier training, and threshold calibration.
Table~\ref{tab:corpus} lists the conditions of each RQ.

\begin{table*}[!t]
\centering
\caption{\corpus conditions: test and development cases, input length percentiles, and first-pass and final yield of the blind-verification gate. Derived conditions are built programmatically from verified text.}
\label{tab:corpus}
\scriptsize
\setlength{\tabcolsep}{8pt}
\setlength{\aboverulesep}{0pt}\setlength{\belowrulesep}{0pt}\setlength{\extrarowheight}{0pt}
\renewcommand{\arraystretch}{0.94}
\def\ebtab{%
\begin{tabular}{llrrrrrrr}
\toprule
\rowcolor{ebBlue}Condition & Kind & \ebh{$N$}{test} & \ebh{$N$}{dev} & \ebh{Length}{p5} & \ebh{Length}{p50} & \ebh{Length}{p95} & \ebh{First-pass}{yield} & \ebh{Final}{yield} \\
\midrule
\rowcolor{ebAmber!45}\multicolumn{9}{l}{\textbf{\textit{RQ1a}}} \\
Base length & derived & 300 & 60 & 24 & 38 & 64 & -- & -- \\
Padded to 512 tokens & derived & 300 & 60 & 510 & 511 & 512 & -- & -- \\
Padded to 2{,}048 tokens & derived & 300 & 60 & 2046 & 2047 & 2048 & -- & -- \\
Padded to 8{,}192 tokens & derived & 300 & 60 & 8190 & 8191 & 8192 & -- & -- \\
Padded to 16{,}384 tokens & derived & 300 & 60 & 16382 & 16383 & 16384 & -- & -- \\
\rowcolor{ebAmber!45}\multicolumn{9}{l}{\textbf{\textit{RQ1b}}} \\
$k=1$ request per message & derived & 300 & 60 & 28 & 42 & 68 & -- & -- \\
$k=2$ requests per message & derived & 300 & 60 & 66 & 87 & 123 & -- & -- \\
$k=4$ requests per message & derived & 300 & 60 & 149 & 177 & 223 & -- & -- \\
$k=8$ requests per message & derived & 300 & 60 & 302.9 & 356 & 432 & -- & -- \\
\rowcolor{ebAmber!45}\multicolumn{9}{l}{\textbf{\textit{RQ2}}} \\
Clean & generated & 300 & 60 & 24 & 38 & 64 & 1.000 & 1.000 \\
Code-switched & generated & 300 & 60 & 30 & 43 & 68.1 & 0.933 & 1.000 \\
Colloquial & generated & 300 & 60 & 25 & 41 & 69 & 1.000 & 1.000 \\
Default-baiting & generated & 300 & 60 & 33 & 44 & 69 & 0.900 & 1.000 \\
Key--value & generated & 300 & 60 & 14 & 27 & 53 & 0.997 & 1.000 \\
Negation-heavy & generated & 300 & 60 & 30 & 45 & 69.1 & 1.000 & 1.000 \\
Noisy & derived & 300 & 60 & 24 & 38 & 66 & -- & -- \\
Self-revising & generated & 300 & 60 & 31 & 46 & 79.1 & 0.994 & 1.000 \\
\rowcolor{ebAmber!45}\multicolumn{9}{l}{\textbf{\textit{RQ3}}} \\
$F=4$, low density & generated & 300 & 60 & 21 & 37.5 & 65 & 0.997 & 1.000 \\
$F=4$, medium density & generated & 300 & 60 & 36 & 51 & 72.1 & 1.000 & 1.000 \\
$F=4$, high density & generated & 300 & 60 & 61 & 77.5 & 108 & 1.000 & 1.000 \\
$F=6$, low density & generated & 300 & 60 & 33 & 50 & 76.1 & 1.000 & 1.000 \\
$F=6$, medium density & generated & 300 & 60 & 46 & 61 & 90 & 0.978 & 1.000 \\
$F=6$, high density & generated & 300 & 60 & 67 & 81 & 108 & 0.983 & 1.000 \\
$F=8$, low density & generated & 300 & 60 & 29 & 41 & 77 & 0.983 & 1.000 \\
$F=8$, medium density & generated & 300 & 60 & 46 & 65 & 101.1 & 0.989 & 1.000 \\
$F=8$, high density & generated & 300 & 60 & 68 & 87 & 124 & 1.000 & 1.000 \\
\rowcolor{ebAmber!45}\multicolumn{9}{l}{\textbf{\textit{RQ4}}} \\
$K=4$ services & generated & 300 & 60 & 27 & 46 & 76 & 0.975 & 1.000 \\
$K=15$ services & generated & 300 & 60 & 33 & 52 & 82.1 & 0.978 & 1.000 \\
$K=64$ services & generated & 300 & 60 & 38 & 55 & 81.1 & 0.997 & 1.000 \\
$K=128$ services & generated & 300 & 60 & 36 & 54.5 & 78 & 0.978 & 1.000 \\
$K=254$ services & generated & 300 & 60 & 34 & 51 & 75.1 & 0.992 & 1.000 \\
25\% churn at $K=64$ & generated & 300 & 60 & 35 & 53 & 81 & 0.989 & 1.000 \\
50\% churn at $K=64$ & generated & 300 & 60 & 35 & 56 & 83.1 & 0.992 & 1.000 \\
\bottomrule
\end{tabular}}
\def\ebh#1#2{#1 #2}\sbox0{\ebtab}
\ifdim\wd0>\dimexpr\linewidth-1pt\relax\def\ebh#1#2{\shortstack{#1\\{}#2}}\sbox0{\ebtab}\fi
\setlength{\tabcolsep}{\dimexpr\tabcolsep+(\linewidth-\wd0-1pt)/18\relax}
\ifdim\tabcolsep<1.5pt\setlength{\tabcolsep}{1.5pt}\fi
\typeout{EBTAB tab_corpus natural=\the\wd0\space line=\the\linewidth\space sep=\the\tabcolsep}
\ebtab
\end{table*}

RQ1 pads each request with unrelated logs, ticket threads, or configuration
dumps up to 16{,}384 tokens (RQ1a), and bundles $k\in\{1,2,4,8\}$ requests into
one message (RQ1b). RQ2 keeps the label tuples of the clean condition and
rewrites their text as colloquial, noisy, negation-heavy, code-switched,
default-baiting, self-revising, or key--value input. RQ3 crosses contract size
$F\in\{4,6,8\}$ with low, medium, and high constraint density. RQ4 varies the
catalog size $K\in\{4,15,64,128,254\}$ over nested catalogs of 254 edge
services with deliberate near neighbors. Its churn conditions replace 25\% or 50\% of a
64-service catalog with services outside it, which the classifier never saw. Unsupported
requests make up 10\% of each RQ4 condition.

\subsection{Service Workloads and Matched Comparisons}
Part~A varies one factor at a time around a center point of $\lambda=4$
requests/s, a 2\,s deadline, five edge nodes, changing text, and caching
disabled. The load sweep covers $\lambda\in\{1,2,4,8,16\}$, the deadline sweep
$D\in\{0.5,1,2,4\}$\,s, and the topology sweep 5 to 40 edge nodes at
$\lambda=8$. A reuse block crosses changing or repeated descriptions with
caching disabled or enabled, and one bursty condition matches the mean rate.
The 15 cells each hold 300 arrivals drawn from the clean four-field test
split. Live admission events feed modeled execution on the same elapsed-time
axis, and no images are processed in Part~A.

Part~B holds the corresponding request descriptions, image assignments,
arrival schedules, and service configuration fixed across all interpreters.
Its eight conditions cross steady or bursty arrivals, changing or repeated
descriptions, and caching disabled or enabled. Complete interpreter runs are
executed sequentially in randomized order. Consequently, matching controls
the workload but does not eliminate variation in provider or network state.

The real service uses Tesseract recognition~\cite{smith2007overview} on images
from the IIIT5K scene-text dataset~\cite{mishra2012scene}. Three workers run
identical container images, whereas link emulation gives the second edge node and the
cloud node their delays and bandwidths. Separate development
images calibrate the scheduler's service-duration estimates. Evaluation
images are selected before recognition results are observed and are not
filtered by whether a tier recognizes them correctly. The tiers are execution
configurations and do not guarantee recognition quality. Each condition
includes 180 OCR requests and 60 requests for counting or detection, which the
testbed does not run and must reject. Changing-text conditions use 180 distinct
OCR descriptions, drawn from the clean and the four-field constraint-density
test splits. Images and
intent descriptions are reused across conditions, while a new service run
is still required after an interpretation-cache hit. Deadlines and offered
loads are controlled study parameters.

\subsection{Correctness and Completion}
\label{sec:metrics}
Exact semantic correctness requires all interpreted fields to equal
their reference values. RQ1--RQ4 measure it on every test case, together with
the valid-output rate and three field-level failure rates. An unsafe decision
permits remote execution where the reference keeps the payload on site. A
spurious specification states a value where the reference is unspecified, and
a missed constraint leaves a stated requirement unspecified. RQ4 adds top-1
service accuracy on seen and unseen services and the F1 score for detecting
unsupported requests.

Operational completion instead requires a supported request to finish
execution by its deadline while satisfying the reference service,
locality, minimum tier, and mapped priority. Its strict companion additionally
requires exact field equality. Part~B's correct completion adds equality
between the returned OCR text and the reference text after canonical Unicode
normalization and removal of surrounding whitespace. Case and internal
characters are retained. These criteria distinguish interpretation
accuracy, execution compliance, and service-output correctness.

For a given backend and condition, let $K$ be the number of supported
requests and let $C$ be the number meeting the applicable completion
criterion. For $K>0$, the completion rate is $C/K$. Because matched conditions share $K$,
their completion counts can also be compared directly. Unsupported requests
and their correct rejections are reported separately. Outcome distributions
over all arrivals retain both unsupported requests and failures. This
separation prevents successful rejections from inflating service completion.

\subsection{Latency and API Cost}
Decision latency is measured from client call initiation to receipt of the
full interpretation response. Admission waiting is accounted for separately
in the service timeline. We report medians and empirical 95th percentiles
(p95), together with the interquartile range (IQR) as jitter and the
probability that a decision exceeds a latency budget $\tau$.

For backend $m$ and request $i$, let $f_i^m$ be the time when its service
response reaches the originating controller. Its full request latency is
$T_i^m=f_i^m-a_i$. The full request latency $T_i^m$ includes admission waiting,
interpretation, service queueing, transfer, and execution. For a matched pair of \jev and an LLM $L$,
let $\mathcal{S}$ contain the requests completed correctly by both backends,
and let $N=|\mathcal{S}|$. The median paired saving
$\Delta_{\mathrm{paired}}$ is
\begin{equation}
\Delta_{\mathrm{paired}}=
\operatorname{median}_{i\in\mathcal{S}}
\left(T_i^{L}-T_i^{\mathrm{Jev}}\right).
\label{eq:paired}
\end{equation}
Positive values favor Jev. The percentage reduction of median latency uses
a different summary. With $M_m$ denoting the median of $T_i^m$ over $\mathcal{S}$,
the percentage reduction $R_T$ is
\begin{equation}
R_T=100\left(1-\frac{M_{\mathrm{Jev}}}{M_{L}}\right).
\label{eq:latency-reduction}
\end{equation}
Equation~\ref{eq:paired} need not equal $M_{L}-M_{\mathrm{Jev}}$.
Because both summaries are conditional on shared success, we report $N$ and
each backend's completion count alongside them.

For a backend and condition, let $A$ be the provider-reported interpretation fees in
United States dollars (USD) and let $c$ be API cost per completion. Using
the corresponding completion count $C$, we define
\begin{equation}
c=\frac{A}{C},\qquad C>0.
\label{eq:cost}
\end{equation}
The numerator includes all interpretation calls for that condition, including
unsupported requests, unsuccessful jobs, and calls ending after their
deadline. It excludes warmups. RQ1--RQ4 use exactly correct decisions for $C$
and report $c$ per 1{,}000 correct decisions. Part~A uses operational
completion and Part~B correct OCR completion.
Cost is undefined when $C=0$. Relative cost reductions use the same ratio
as Equation~\ref{eq:latency-reduction}, with $c$ replacing $M$.
These fees exclude computation, communication infrastructure, energy, and
maintenance. Because self-hosted models carry no API fee, we report energy
per decision for them, integrated from the accelerator power trace.

\subsection{Statistical Analysis}
Hypotheses H1--H8 and their decision rules were fixed before the runs. Within
each RQ, the unit is the test case, and interpreters are compared on the same
cases. Confidence intervals come from 10{,}000 case bootstrap resamples. Paired
proportions use exact McNemar tests, latencies Wilcoxon signed-rank tests,
and the six-way comparison of unsafe decisions Cochran's $Q$. Holm's method
controls the error rate within each hypothesis family, and a contrast counts
as resolved when its 95\% interval excludes the null value and its adjusted
$p$ is below 0.05. Latency contrasts are confirmatory only within a deployment
class, hosted against hosted and self-hosted against self-hosted. Because RQ5
arrivals share queues and caches, its intervals use a block bootstrap over 30
contiguous arrivals. Cells in which an interpreter returns fewer than half
valid outputs are still tested and are flagged. Eight recorded deviations from the
protocol include the \glm reasoning setting, the 15-service catalog level chosen
for \semif's option limit, a single run per case, and the corpus generators and
verifier. The remaining deviations cover the accelerator for the self-hosted models, shared reading rules in
the field instructions, one availability re-run for \glm, and the classifier
training set.

\subsection{Implementation and Settings}
\label{sec:settings}
\smallskip\noindent\textbf{Interpretation backends. }The interpretation-layer runs were executed on September 24--25, 2026. Jev
uses OpenRouter's Decisions endpoint with identifier typesafe/jev-1.13.
Responses identify jev-1.13-20260917 and provider TypeSafe. One native Choice
question per contract field returns the intent fields in one request.
\deepseek and \glm are pinned to Together and \qwenflash to Alibaba, its only
endpoint. All three have provider fallback disabled and data collection denied. They
return a strict JSON object with temperature zero. Reasoning is disabled and
recorded reasoning-token counts are zero for \deepseek and \qwenflash. Because no \glm
endpoint accepts disabled reasoning, the model runs at the lowest setting. Its
reasoning tokens count toward its output tokens, latency, and cost. Calls have
a 30\,s timeout, no retries, and no caching.

\semif (SemIf-OpenJev commit 23cf1f39 over Qwen/Qwen3.5-4B revision
851bf6e8), \laya (revision bc76315b), and \qwenjson (the same Qwen3.5-4B
weights with the LLM prompt and greedy decoding) run one at a time on an
NVIDIA H100 NVL. These models use PyTorch 2.13.0 and Transformers 5.17.0 and
execute in time blocks separate from the hosted runs. The fixed rule parser uses keywords and regular
expressions without tuning on evaluation wording. The \reranker scores the
cosine similarity between a request and each service description and abstains
below a threshold of 0.3691, calibrated on the development split. The \clf
heads are fine-tuned on the development split plus one description example per
service. The fine-tuning fixes a maximum length of 128 tokens, learning rate
$5\times10^{-5}$, batch size 16, ten epochs, and seed 20260924 without a search.

\smallskip\noindent\textbf{Admission and execution controls. }Both parts of RQ5 use four concurrent interpretation slots and an admission
queue of capacity 32, with no automatic retries or repair calls. Connections
are warmed before measurement. Each condition starts with an empty cache keyed
by normalized full request text and extraction policy. The controller checks
it before queueing and after obtaining a slot, but concurrent misses are not
coalesced. Self-hosted interpreters run on the same host as the controller,
in their own time blocks.

Each worker has one nonpreemptive queue. Urgent and ordinary jobs have
priority values zero and one, respectively, and equal-priority jobs follow
enqueue order. The controller verifies the returned node, tier, priority, and image
identity. A timed-out OCR process is terminated before capacity is released.

\smallskip\noindent\textbf{Part~A parameters. }Steady traffic follows a Poisson process. Bursty traffic alternates 0.5 and
8 requests/s in 20\,s segments. The execution model has $N$ edge nodes and one
cloud node, with payloads of 0.25--2\,MB. Local, other-edge, and cloud links
have propagation delays of 2, 20, and 60\,ms and bandwidths of 1000, 100, and
50\,Mbit/s, respectively. Transfer over these links also includes
serialization. Base durations
are 40\,ms for counting, 80\,ms for detection, and 60\,ms for OCR. The high tier
multiplies these by 1.8. Edge node factors are spread over 1.0--1.3, and the
cloud factor is 0.65. Each node has one priority worker. Repeated-text cells
use eight recurring descriptions.

\smallskip\noindent\textbf{Part~B parameters and scoring. }The three workers run the same container image with one Tesseract build and
English recognition in single-word mode. Standard and high tiers use
tessdata\_fast and tessdata\_best, respectively, with identical weights across
nodes. The second edge node's link is emulated at 10\,ms one-way delay and
100\,Mbit/s, and the cloud node's at 30\,ms and 50\,Mbit/s. A seeded split of
IIIT5K identifiers selects calibration and test images before recognition
outcomes are known. Median request--response durations on 40 calibration
images are 52 and 59\,ms on the local node for the standard and high tiers. The
corresponding durations are 76 and 89\,ms on the second edge node and 128 and
137\,ms on the cloud node.
Steady arrivals average 2 requests/s, and bursty arrivals follow the Part~A
pattern. Repeated-text conditions cycle six OCR and two unsupported
descriptions. The request deadline is 2\,s, and each run holds 240 arrivals.

OCR comparison uses case-preserving character annotations without a
recognition lexicon, normalizes Unicode to Normalization Form C (NFC), and
strips surrounding whitespace. Each request is evaluated on its recorded
timing and output, and cached decisions do not replace service execution.

\section{Results}
\label{sec:results}

Table~\ref{tab:hypotheses} summarizes every pre-registered contrast of H1--H5,
with its estimate, 95\% confidence interval, and Holm-adjusted $p$. The
subsections below examine these contrasts per RQ. H6--H8 are tested in
Section~\ref{sec:rq5} under the same criterion, with the Holm adjustment
applied within each part.

\subsection{RQ1: Input Scale}
\label{sec:rq1}
\begin{figure*}[!t]
\centering
\includegraphics[width=7.00in]{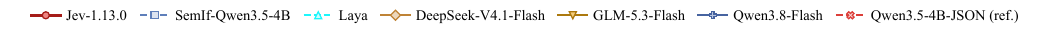}\\[-2pt]
\subfloat[p50 latency vs.\ input length]{\includegraphics[width=1.70in]{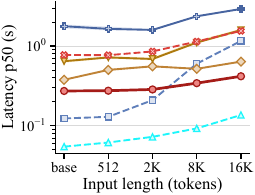}\label{fig:rq1_a}}\hfill
\subfloat[EM vs.\ input length]{\includegraphics[width=1.70in]{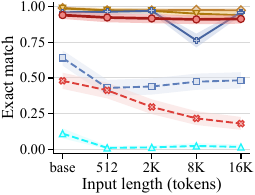}\label{fig:rq1_b}}\hfill
\subfloat[Per-request p50 vs.\ $k$]{\includegraphics[width=1.70in]{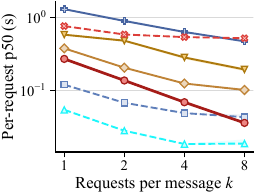}\label{fig:rq1_c}}\hfill
\subfloat[Request-level EM vs.\ $k$]{\includegraphics[width=1.70in]{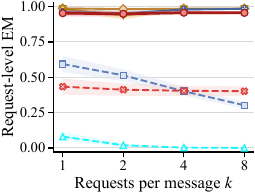}\label{fig:rq1_d}}
\caption{RQ1, input scale. (a) Median decision latency and (b) exact match (EM) as irrelevant context pads each request to 16{,}384 tokens. (c) Per-request median latency and (d) request-level EM when $k$ requests share one message. Bands are 95\% bootstrap confidence intervals; latency axes are logarithmic.}
\label{fig:rq1}
\end{figure*}

\begin{table*}[!t]
\centering
\caption{RQ1a, input length. EM with 95\% confidence interval, valid-output rate, macro field accuracy, decision latency percentiles, output tokens, API fees per 1{,}000 correct decisions, and energy per decision. Best value per block in bold.}
\label{tab:rq1a}
\scriptsize
\setlength{\tabcolsep}{3pt}
\setlength{\aboverulesep}{0pt}\setlength{\belowrulesep}{0pt}\setlength{\extrarowheight}{0pt}
\renewcommand{\arraystretch}{0.94}
\def\ebtab{%
\begin{tabular}{lrrrrrrrrr}
\toprule
\rowcolor{ebBlue}Model & \ebh{EM [95\% CI]}{$\uparrow$} & \ebh{Valid}{$\uparrow$} & \ebh{Macro acc.}{$\uparrow$} & \ebh{p50}{(s) $\downarrow$} & \ebh{p95}{(s) $\downarrow$} & \ebh{p99}{(s) $\downarrow$} & \ebh{Output}{tokens} & \ebh{USD per 1k}{correct $\downarrow$} & \ebh{Energy}{(J/dec.) $\downarrow$} \\
\midrule
\rowcolor{ebAmber!45}\multicolumn{10}{l}{\textbf{\textit{Base length}}} \\
\rowcolor{ebCoral!10}{\jev} & 0.940\,{\color{black!60}[0.913, 0.967]} & 1.000 & 0.985 & 0.274 & 0.381 & 0.536 & 181 & \cellcolor{ebCoral!35}\textbf{0.03650} & -- \\
{\semif} & 0.640\,{\color{black!60}[0.587, 0.693]} & 1.000 & 0.902 & 0.124 & 0.296 & 0.323 & 0 & -- & 23.7 \\
{\laya} & 0.110\,{\color{black!60}[0.077, 0.147]} & 1.000 & 0.671 & \cellcolor{ebCoral!35}\textbf{0.055} & \cellcolor{ebCoral!35}\textbf{0.222} & \cellcolor{ebCoral!35}\textbf{0.242} & 0 & -- & \cellcolor{ebCoral!35}\textbf{8.4} \\
{\deepseek} & \cellcolor{ebCoral!35}\textbf{0.990}\,{\color{black!60}[0.977, 1.000]} & 1.000 & \cellcolor{ebCoral!35}\textbf{0.998} & 0.378 & 1.054 & 4.933 & 26 & 0.09051 & -- \\
{\glm} & 0.987\,{\color{black!60}[0.973, 0.997]} & 1.000 & 0.997 & 0.646 & 2.685 & 3.363 & 36 & 0.04048 & -- \\
{\qwenflash} & 0.960\,{\color{black!60}[0.937, 0.980]} & 0.997 & 0.988 & 1.774 & 3.668 & 5.774 & 30 & 0.03797 & -- \\
{\qwenjson{} (ref.)} & 0.480\,{\color{black!60}[0.423, 0.537]} & 1.000 & 0.854 & 0.774 & 0.868 & 1.036 & 30 & -- & 127.2 \\
\rowcolor{ebAmber!45}\multicolumn{10}{l}{\textbf{\textit{Padded to 512 tokens}}} \\
\rowcolor{ebCoral!10}{\jev} & 0.923\,{\color{black!60}[0.890, 0.953]} & 1.000 & 0.981 & 0.276 & 0.359 & 0.505 & 181 & \cellcolor{ebCoral!35}\textbf{0.06585} & -- \\
{\semif} & 0.430\,{\color{black!60}[0.373, 0.487]} & 1.000 & 0.805 & 0.130 & 0.231 & 0.290 & 0 & -- & 30.9 \\
{\laya} & 0.010\,{\color{black!60}[0.000, 0.023]} & 1.000 & 0.525 & \cellcolor{ebCoral!35}\textbf{0.062} & \cellcolor{ebCoral!35}\textbf{0.150} & \cellcolor{ebCoral!35}\textbf{0.191} & 0 & -- & \cellcolor{ebCoral!35}\textbf{11.6} \\
{\deepseek} & \cellcolor{ebCoral!35}\textbf{0.973}\,{\color{black!60}[0.953, 0.990]} & 1.000 & \cellcolor{ebCoral!35}\textbf{0.993} & 0.503 & 0.883 & 2.113 & 25 & 0.2534 & -- \\
{\glm} & \cellcolor{ebCoral!35}\textbf{0.973}\,{\color{black!60}[0.953, 0.990]} & 1.000 & \cellcolor{ebCoral!35}\textbf{0.993} & 0.720 & 2.469 & 9.620 & 26 & 0.1146 & -- \\
{\qwenflash} & 0.963\,{\color{black!60}[0.940, 0.983]} & 0.997 & 0.988 & 1.643 & 4.566 & 7.140 & 36 & 0.1462 & -- \\
{\qwenjson{} (ref.)} & 0.413\,{\color{black!60}[0.357, 0.467]} & 1.000 & 0.818 & 0.768 & 0.872 & 1.218 & 30 & -- & 130.5 \\
\rowcolor{ebAmber!45}\multicolumn{10}{l}{\textbf{\textit{Padded to 2{,}048 tokens}}} \\
\rowcolor{ebCoral!10}{\jev} & 0.917\,{\color{black!60}[0.883, 0.947]} & 1.000 & 0.979 & 0.286 & 0.404 & 0.522 & 181 & \cellcolor{ebCoral!35}\textbf{0.1606} & -- \\
{\semif} & 0.440\,{\color{black!60}[0.383, 0.497]} & 1.000 & 0.819 & 0.210 & 0.334 & 0.359 & 0 & -- & 61.6 \\
{\laya} & 0.013\,{\color{black!60}[0.003, 0.027]} & 1.000 & 0.380 & \cellcolor{ebCoral!35}\textbf{0.073} & \cellcolor{ebCoral!35}\textbf{0.074} & \cellcolor{ebCoral!35}\textbf{0.075} & 0 & -- & \cellcolor{ebCoral!35}\textbf{14.4} \\
{\deepseek} & \cellcolor{ebCoral!35}\textbf{0.973}\,{\color{black!60}[0.953, 0.990]} & 1.000 & \cellcolor{ebCoral!35}\textbf{0.993} & 0.557 & 1.076 & 19.198 & 25 & 0.7535 & -- \\
{\glm} & 0.970\,{\color{black!60}[0.950, 0.987]} & 1.000 & 0.992 & 0.689 & 2.518 & 3.177 & 26 & 0.3642 & -- \\
{\qwenflash} & 0.970\,{\color{black!60}[0.950, 0.987]} & 1.000 & \cellcolor{ebCoral!35}\textbf{0.993} & 1.594 & 2.408 & 4.261 & 40 & 0.4526 & -- \\
{\qwenjson{} (ref.)} & 0.297\,{\color{black!60}[0.247, 0.350]} & 1.000 & 0.765 & 0.857 & 0.981 & 1.545 & 31 & -- & 162.0 \\
\rowcolor{ebAmber!45}\multicolumn{10}{l}{\textbf{\textit{Padded to 8{,}192 tokens}}} \\
\rowcolor{ebCoral!10}{\jev} & 0.910\,{\color{black!60}[0.877, 0.940]} & 1.000 & 0.977 & 0.343 & 0.525 & 0.839 & 181 & \cellcolor{ebCoral!35}\textbf{0.5417} & -- \\
{\semif} & 0.473\,{\color{black!60}[0.417, 0.530]} & 1.000 & 0.823 & 0.602 & 0.718 & 0.748 & 0 & -- & 194.9 \\
{\laya} & 0.023\,{\color{black!60}[0.007, 0.043]} & 1.000 & 0.395 & \cellcolor{ebCoral!35}\textbf{0.093} & \cellcolor{ebCoral!35}\textbf{0.094} & \cellcolor{ebCoral!35}\textbf{0.095} & 0 & -- & \cellcolor{ebCoral!35}\textbf{16.3} \\
{\deepseek} & \cellcolor{ebCoral!35}\textbf{0.977}\,{\color{black!60}[0.957, 0.993]} & 1.000 & \cellcolor{ebCoral!35}\textbf{0.993} & 0.519 & 0.785 & 1.232 & 25 & 2.753 & -- \\
{\glm} & 0.950\,{\color{black!60}[0.923, 0.973]} & 0.993 & 0.981 & 1.107 & 5.631 & 29.708 & 26 & 1.384 & -- \\
{\qwenflash} & 0.763\,{\color{black!60}[0.713, 0.810]} & 1.000 & 0.909 & 2.364 & 4.791 & 8.870 & 38 & 2.144 & -- \\
{\qwenjson{} (ref.)} & 0.217\,{\color{black!60}[0.170, 0.267]} & 1.000 & 0.714 & 1.140 & 1.206 & 1.859 & 31 & -- & 279.8 \\
\rowcolor{ebAmber!45}\multicolumn{10}{l}{\textbf{\textit{Padded to 16{,}384 tokens}}} \\
\rowcolor{ebCoral!10}{\jev} & 0.913\,{\color{black!60}[0.880, 0.943]} & 1.000 & 0.978 & 0.418 & 0.523 & 1.198 & 181 & \cellcolor{ebCoral!35}\textbf{1.044} & -- \\
{\semif} & 0.483\,{\color{black!60}[0.427, 0.540]} & 1.000 & 0.829 & 1.159 & 1.238 & 1.288 & 0 & -- & 378.8 \\
{\laya} & 0.017\,{\color{black!60}[0.003, 0.033]} & 1.000 & 0.392 & \cellcolor{ebCoral!35}\textbf{0.137} & \cellcolor{ebCoral!35}\textbf{0.140} & \cellcolor{ebCoral!35}\textbf{0.142} & 0 & -- & \cellcolor{ebCoral!35}\textbf{20.4} \\
{\deepseek} & \cellcolor{ebCoral!35}\textbf{0.973}\,{\color{black!60}[0.953, 0.990]} & 1.000 & \cellcolor{ebCoral!35}\textbf{0.993} & 0.638 & 1.023 & 1.183 & 25 & 5.455 & -- \\
{\glm} & 0.940\,{\color{black!60}[0.913, 0.967]} & 0.980 & 0.968 & 1.588 & 3.939 & 12.754 & 26 & 2.727 & -- \\
{\qwenflash} & 0.960\,{\color{black!60}[0.937, 0.980]} & 1.000 & 0.990 & 2.920 & 4.274 & 5.400 & 38 & 3.373 & -- \\
{\qwenjson{} (ref.)} & 0.180\,{\color{black!60}[0.137, 0.227]} & 1.000 & 0.694 & 1.550 & 1.619 & 2.384 & 31 & -- & 444.7 \\
\bottomrule
\end{tabular}}
\def\ebh#1#2{#1 #2}\sbox0{\ebtab}
\ifdim\wd0>\dimexpr\linewidth-1pt\relax\def\ebh#1#2{\shortstack{#1\\{}#2}}\sbox0{\ebtab}\fi
\setlength{\tabcolsep}{\dimexpr\tabcolsep+(\linewidth-\wd0-1pt)/20\relax}
\ifdim\tabcolsep<1.5pt\setlength{\tabcolsep}{1.5pt}\fi
\typeout{EBTAB tab_rq1a natural=\the\wd0\space line=\the\linewidth\space sep=\the\tabcolsep}
\ebtab
\end{table*}

\begin{table*}[!t]
\centering
\caption{RQ1b, bundled requests. Request-level EM, share of messages with every request correct, per-request and per-message latency, API fees, and energy for $k$ requests per message.}
\label{tab:rq1b}
\scriptsize
\setlength{\tabcolsep}{5pt}
\setlength{\aboverulesep}{0pt}\setlength{\belowrulesep}{0pt}\setlength{\extrarowheight}{0pt}
\renewcommand{\arraystretch}{0.94}
\def\ebtab{%
\begin{tabular}{lrrrrrrr}
\toprule
\rowcolor{ebBlue}Model & \ebh{Request EM [95\% CI]}{$\uparrow$} & \ebh{Message}{all-correct $\uparrow$} & \ebh{Per-request}{p50 (s) $\downarrow$} & \ebh{Message}{p50 (s) $\downarrow$} & \ebh{Message}{p95 (s) $\downarrow$} & \ebh{USD per 1k}{correct $\downarrow$} & \ebh{Energy}{(J/dec.) $\downarrow$} \\
\midrule
\rowcolor{ebAmber!45}\multicolumn{8}{l}{\textbf{\textit{$k=1$ request per message}}} \\
\rowcolor{ebCoral!10}{\jev} & 0.953\,{\color{black!60}[0.930, 0.977]} & 0.953 & 0.272 & 0.272 & 0.372 & \cellcolor{ebCoral!35}\textbf{0.03617} & -- \\
{\semif} & 0.593\,{\color{black!60}[0.537, 0.650]} & 0.593 & 0.121 & 0.121 & 0.121 & -- & 22.3 \\
{\laya} & 0.080\,{\color{black!60}[0.050, 0.113]} & 0.080 & \cellcolor{ebCoral!35}\textbf{0.055} & \cellcolor{ebCoral!35}\textbf{0.055} & \cellcolor{ebCoral!35}\textbf{0.056} & -- & \cellcolor{ebCoral!35}\textbf{6.6} \\
{\deepseek} & \cellcolor{ebCoral!35}\textbf{0.987}\,{\color{black!60}[0.973, 0.997]} & \cellcolor{ebCoral!35}\textbf{0.987} & 0.379 & 0.379 & 0.552 & 0.1113 & -- \\
{\glm} & 0.983\,{\color{black!60}[0.967, 0.997]} & 0.983 & 0.583 & 0.583 & 0.817 & 0.03997 & -- \\
{\qwenflash} & 0.967\,{\color{black!60}[0.943, 0.987]} & 0.967 & 1.302 & 1.302 & 2.221 & 0.05417 & -- \\
{\qwenjson{} (ref.)} & 0.433\,{\color{black!60}[0.377, 0.490]} & 0.433 & 0.765 & 0.765 & 0.794 & -- & 121.8 \\
\rowcolor{ebAmber!45}\multicolumn{8}{l}{\textbf{\textit{$k=2$ requests per message}}} \\
\rowcolor{ebCoral!10}{\jev} & 0.948\,{\color{black!60}[0.930, 0.965]} & 0.897 & 0.138 & 0.276 & 0.359 & \cellcolor{ebCoral!35}\textbf{0.06665} & -- \\
{\semif} & 0.513\,{\color{black!60}[0.472, 0.555]} & 0.283 & 0.068 & 0.137 & 0.312 & -- & 34.0 \\
{\laya} & 0.018\,{\color{black!60}[0.008, 0.030]} & 0.000 & \cellcolor{ebCoral!35}\textbf{0.028} & \cellcolor{ebCoral!35}\textbf{0.057} & \cellcolor{ebCoral!35}\textbf{0.207} & -- & \cellcolor{ebCoral!35}\textbf{10.5} \\
{\deepseek} & \cellcolor{ebCoral!35}\textbf{0.982}\,{\color{black!60}[0.970, 0.992]} & \cellcolor{ebCoral!35}\textbf{0.963} & 0.206 & 0.413 & 0.626 & 0.1757 & -- \\
{\glm} & 0.937\,{\color{black!60}[0.910, 0.960]} & 0.920 & 0.483 & 0.966 & 4.237 & 0.07183 & -- \\
{\qwenflash} & 0.952\,{\color{black!60}[0.935, 0.968]} & 0.903 & 0.898 & 1.796 & 2.952 & 0.09168 & -- \\
{\qwenjson{} (ref.)} & 0.412\,{\color{black!60}[0.368, 0.455]} & 0.223 & 0.587 & 1.173 & 1.277 & -- & 194.4 \\
\rowcolor{ebAmber!45}\multicolumn{8}{l}{\textbf{\textit{$k=4$ requests per message}}} \\
\rowcolor{ebCoral!10}{\jev} & 0.957\,{\color{black!60}[0.944, 0.968]} & 0.850 & 0.070 & 0.278 & 0.358 & 0.1278 & -- \\
{\semif} & 0.401\,{\color{black!60}[0.369, 0.433]} & 0.047 & 0.049 & 0.198 & 0.330 & -- & 57.1 \\
{\laya} & 0.002\,{\color{black!60}[0.000, 0.004]} & 0.000 & \cellcolor{ebCoral!35}\textbf{0.019} & \cellcolor{ebCoral!35}\textbf{0.075} & \cellcolor{ebCoral!35}\textbf{0.197} & -- & \cellcolor{ebCoral!35}\textbf{17.2} \\
{\deepseek} & \cellcolor{ebCoral!35}\textbf{0.985}\,{\color{black!60}[0.978, 0.992]} & \cellcolor{ebCoral!35}\textbf{0.940} & 0.125 & 0.502 & 0.687 & 0.2669 & -- \\
{\glm} & 0.983\,{\color{black!60}[0.974, 0.990]} & 0.933 & 0.285 & 1.140 & 2.113 & \cellcolor{ebCoral!35}\textbf{0.1207} & -- \\
{\qwenflash} & 0.979\,{\color{black!60}[0.969, 0.988]} & 0.927 & 0.635 & 2.540 & 3.445 & 0.1365 & -- \\
{\qwenjson{} (ref.)} & 0.404\,{\color{black!60}[0.368, 0.440]} & 0.053 & 0.541 & 2.162 & 2.286 & -- & 358.3 \\
\rowcolor{ebAmber!45}\multicolumn{8}{l}{\textbf{\textit{$k=8$ requests per message}}} \\
\rowcolor{ebCoral!10}{\jev} & 0.955\,{\color{black!60}[0.946, 0.964]} & 0.717 & 0.036 & 0.290 & 0.375 & 0.2881 & -- \\
{\semif} & 0.301\,{\color{black!60}[0.278, 0.325]} & 0.003 & 0.044 & 0.348 & 0.773 & -- & 109.6 \\
{\laya} & 0.000\,{\color{black!60}[0.000, 0.000]} & 0.000 & \cellcolor{ebCoral!35}\textbf{0.019} & \cellcolor{ebCoral!35}\textbf{0.152} & \cellcolor{ebCoral!35}\textbf{0.225} & -- & \cellcolor{ebCoral!35}\textbf{40.7} \\
{\deepseek} & \cellcolor{ebCoral!35}\textbf{0.985}\,{\color{black!60}[0.980, 0.990]} & \cellcolor{ebCoral!35}\textbf{0.887} & 0.102 & 0.819 & 1.720 & 0.4706 & -- \\
{\glm} & 0.983\,{\color{black!60}[0.978, 0.988]} & 0.867 & 0.195 & 1.560 & 2.884 & \cellcolor{ebCoral!35}\textbf{0.2264} & -- \\
{\qwenflash} & 0.984\,{\color{black!60}[0.978, 0.989]} & 0.880 & 0.473 & 3.785 & 5.499 & 0.2288 & -- \\
{\qwenjson{} (ref.)} & 0.401\,{\color{black!60}[0.366, 0.435]} & 0.020 & 0.522 & 4.174 & 6.142 & -- & 731.9 \\
\bottomrule
\end{tabular}}
\def\ebh#1#2{#1 #2}\sbox0{\ebtab}
\ifdim\wd0>\dimexpr\linewidth-1pt\relax\def\ebh#1#2{\shortstack{#1\\{}#2}}\sbox0{\ebtab}\fi
\setlength{\tabcolsep}{\dimexpr\tabcolsep+(\linewidth-\wd0-1pt)/16\relax}
\ifdim\tabcolsep<1.5pt\setlength{\tabcolsep}{1.5pt}\fi
\typeout{EBTAB tab_rq1b natural=\the\wd0\space line=\the\linewidth\space sep=\the\tabcolsep}
\ebtab
\end{table*}

Longer inputs slow every interpreter, and \jev least. Its median decision
latency grows from 0.274\,s at the base length to 0.418\,s at 16{,}384 tokens
(Fig.~\ref{fig:rq1} and Table~\ref{tab:rq1a}). Over the same range, \deepseek
grows from 0.378 to 0.638\,s, \glm from 0.646 to 1.588\,s, and \qwenflash from
1.774 to 2.920\,s. \jev is faster than each LLM at every length, and all 15
hosted contrasts resolve (H1). Its growth from base to 16{,}384 tokens is 0.90,
0.62, and 0.93 times that of \deepseek, \glm, and \qwenflash, respectively.
Exact match stays at 0.910--0.940 for \jev and 0.973--0.990 for \deepseek,
whereas \qwenflash drops to 0.763 at 8{,}192 tokens.

The self-hosted pair shows the opposite growth pattern. \semif answers
0.39--0.65\,s faster than \qwenjson at every length, yet its latency grows from
0.124 to 1.159\,s, 4.68 times the growth of \qwenjson. At these weights, input
length costs the readout model more than the generator.

Bundling changes the picture more sharply. Placing eight requests in one
message (Table~\ref{tab:rq1b}) raises \jev's message latency from 0.272 to 0.290\,s, while \deepseek
rises from 0.379 to 0.819\,s, \glm from 0.583 to 1.560\,s, and \qwenflash from
1.302 to 3.785\,s. \jev's per-request latency ratio from one to eight requests
is 0.49, 0.40, and 0.37 times the LLMs' ratios. All three contrasts resolve
(H2), as does the self-hosted contrast (0.53). Request-level exact match stays
at 0.948--0.957 for \jev and 0.982--0.987 for \deepseek, while \semif falls
from 0.593 to 0.301 and \laya from 0.080 to zero.

\takeaway{Input scale}{Irrelevant context and bundled requests leave \jev's
decision time within 0.27--0.42\,s, while every hosted LLM's time at least
doubles with eight bundled requests. The price is 2.8--6.7 points of exact
match relative to \deepseek.}

\subsection{RQ2: Input Quality}
\label{sec:rq2}
\begin{figure*}[!t]
\centering
\includegraphics[width=7.00in]{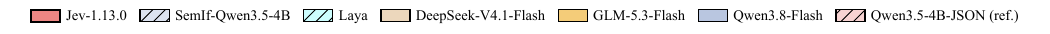}\\[-2pt]
\subfloat[EM]{\includegraphics[width=1.70in]{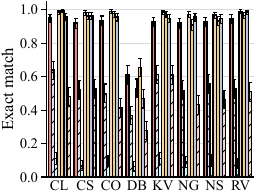}\label{fig:rq2_a}}\hfill
\subfloat[Unsafe-locality rate]{\includegraphics[width=1.70in]{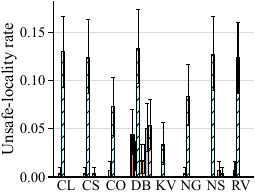}\label{fig:rq2_b}}\hfill
\subfloat[Spurious-specification rate]{\includegraphics[width=1.70in]{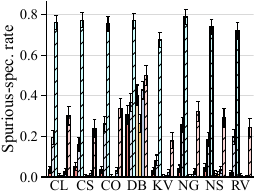}\label{fig:rq2_c}}\hfill
\subfloat[Latency, clean (p50 to p95)]{\includegraphics[width=1.70in]{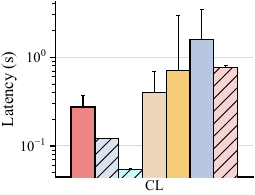}\label{fig:rq2_d}}
\caption{RQ2, input quality, across eight wording conditions that keep the same label tuples. (a) EM, (b) unsafe-locality rate, (c) spurious-specification rate, with 95\% confidence intervals; (d) decision latency on clean requests, bars at the median with whiskers to p95. Conditions: CL clean, CS code-switched, CO colloquial, DB default-baiting, KV key--value, NG negation-heavy, NS noisy, RV self-revising.}
\label{fig:rq2}
\end{figure*}

\begin{table*}[!t]
\centering
\caption{RQ2, input quality. Quality, safety, and latency of every interpreter under each wording condition.}
\label{tab:rq2}
\scriptsize
\setlength{\tabcolsep}{5pt}
\setlength{\aboverulesep}{0pt}\setlength{\belowrulesep}{0pt}\setlength{\extrarowheight}{0pt}
\renewcommand{\arraystretch}{0.94}
\def\ebtab{%
\begin{tabular}{lrrrrrrrr}
\toprule
\rowcolor{ebBlue}Model & \ebh{EM [95\% CI]}{$\uparrow$} & \ebh{Macro acc.}{$\uparrow$} & \ebh{Valid}{$\uparrow$} & \ebh{Unsafe}{$\downarrow$} & \ebh{Spurious}{$\downarrow$} & \ebh{Missed}{$\downarrow$} & \ebh{p50}{(s) $\downarrow$} & \ebh{p95}{(s) $\downarrow$} \\
\midrule
\rowcolor{ebAmber!45}\multicolumn{9}{l}{\textbf{\textit{Clean}}} \\
\rowcolor{ebCoral!10}{\jev} & 0.950\,{\color{black!60}[0.923, 0.973]} & 0.988 & 1.000 & 0.000 & 0.033 & \cellcolor{ebCoral!35}\textbf{0.001} & 0.273 & 0.370 \\
{\semif} & 0.640\,{\color{black!60}[0.583, 0.693]} & 0.902 & 1.000 & 0.003 & 0.195 & 0.031 & 0.122 & 0.123 \\
{\laya} & 0.110\,{\color{black!60}[0.077, 0.147]} & 0.671 & 1.000 & 0.130 & 0.762 & 0.005 & \cellcolor{ebCoral!35}\textbf{0.055} & \cellcolor{ebCoral!35}\textbf{0.056} \\
{\deepseek} & 0.987\,{\color{black!60}[0.973, 0.997]} & 0.997 & 1.000 & 0.000 & 0.007 & \cellcolor{ebCoral!35}\textbf{0.001} & 0.398 & 0.692 \\
{\glm} & \cellcolor{ebCoral!35}\textbf{0.993}\,{\color{black!60}[0.983, 1.000]} & \cellcolor{ebCoral!35}\textbf{0.998} & 1.000 & 0.000 & \cellcolor{ebCoral!35}\textbf{0.002} & \cellcolor{ebCoral!35}\textbf{0.001} & 0.706 & 2.940 \\
{\qwenflash} & 0.960\,{\color{black!60}[0.937, 0.980]} & 0.988 & 0.997 & 0.000 & 0.026 & 0.004 & 1.588 & 3.450 \\
{\qwenjson{} (ref.)} & 0.480\,{\color{black!60}[0.423, 0.537]} & 0.854 & 1.000 & 0.000 & 0.301 & 0.026 & 0.764 & 0.811 \\
\rowcolor{ebAmber!45}\multicolumn{9}{l}{\textbf{\textit{Code-switched}}} \\
\rowcolor{ebCoral!10}{\jev} & 0.920\,{\color{black!60}[0.890, 0.950]} & 0.980 & 1.000 & 0.000 & 0.049 & \cellcolor{ebCoral!35}\textbf{0.003} & 0.278 & 0.366 \\
{\semif} & 0.523\,{\color{black!60}[0.467, 0.580]} & 0.865 & 0.997 & 0.003 & 0.160 & 0.076 & 0.121 & 0.122 \\
{\laya} & 0.067\,{\color{black!60}[0.040, 0.097]} & 0.605 & 1.000 & 0.123 & 0.772 & 0.014 & \cellcolor{ebCoral!35}\textbf{0.055} & \cellcolor{ebCoral!35}\textbf{0.056} \\
{\deepseek} & \cellcolor{ebCoral!35}\textbf{0.983}\,{\color{black!60}[0.967, 0.997]} & \cellcolor{ebCoral!35}\textbf{0.996} & 1.000 & 0.000 & 0.005 & 0.004 & 0.490 & 1.213 \\
{\glm} & 0.963\,{\color{black!60}[0.940, 0.983]} & 0.991 & 1.000 & 0.003 & \cellcolor{ebCoral!35}\textbf{0.002} & 0.010 & 0.729 & 3.115 \\
{\qwenflash} & 0.963\,{\color{black!60}[0.940, 0.983]} & 0.991 & 1.000 & 0.000 & 0.019 & \cellcolor{ebCoral!35}\textbf{0.003} & 1.416 & 2.726 \\
{\qwenjson{} (ref.)} & 0.527\,{\color{black!60}[0.470, 0.583]} & 0.859 & 1.000 & 0.000 & 0.238 & 0.027 & 0.765 & 0.813 \\
\rowcolor{ebAmber!45}\multicolumn{9}{l}{\textbf{\textit{Colloquial}}} \\
\rowcolor{ebCoral!10}{\jev} & 0.937\,{\color{black!60}[0.907, 0.963]} & 0.983 & 1.000 & 0.000 & 0.033 & 0.008 & 0.268 & 0.372 \\
{\semif} & 0.500\,{\color{black!60}[0.443, 0.557]} & 0.853 & 1.000 & 0.007 & 0.261 & 0.043 & 0.121 & 0.122 \\
{\laya} & 0.097\,{\color{black!60}[0.063, 0.130]} & 0.656 & 1.000 & 0.073 & 0.755 & 0.009 & \cellcolor{ebCoral!35}\textbf{0.055} & \cellcolor{ebCoral!35}\textbf{0.055} \\
{\deepseek} & \cellcolor{ebCoral!35}\textbf{0.990}\,{\color{black!60}[0.977, 1.000]} & \cellcolor{ebCoral!35}\textbf{0.998} & 1.000 & 0.000 & 0.005 & 0.001 & 0.392 & 0.629 \\
{\glm} & 0.977\,{\color{black!60}[0.957, 0.990]} & 0.994 & 1.000 & 0.000 & \cellcolor{ebCoral!35}\textbf{0.000} & 0.009 & 0.739 & 1.748 \\
{\qwenflash} & 0.957\,{\color{black!60}[0.933, 0.977]} & 0.989 & 1.000 & 0.000 & 0.031 & \cellcolor{ebCoral!35}\textbf{0.000} & 1.790 & 4.053 \\
{\qwenjson{} (ref.)} & 0.413\,{\color{black!60}[0.357, 0.470]} & 0.833 & 1.000 & 0.000 & 0.339 & 0.035 & 0.763 & 0.809 \\
\rowcolor{ebAmber!45}\multicolumn{9}{l}{\textbf{\textit{Default-baiting}}} \\
\rowcolor{ebCoral!10}{\jev} & 0.610\,{\color{black!60}[0.553, 0.667]} & 0.885 & 1.000 & 0.043 & 0.306 & 0.009 & 0.262 & 0.351 \\
{\semif} & 0.367\,{\color{black!60}[0.313, 0.423]} & 0.792 & 1.000 & 0.020 & 0.367 & 0.068 & 0.122 & 0.122 \\
{\laya} & 0.063\,{\color{black!60}[0.037, 0.093]} & 0.612 & 1.000 & 0.133 & 0.769 & 0.021 & \cellcolor{ebCoral!35}\textbf{0.055} & \cellcolor{ebCoral!35}\textbf{0.056} \\
{\deepseek} & 0.530\,{\color{black!60}[0.473, 0.587]} & 0.858 & 1.000 & \cellcolor{ebCoral!35}\textbf{0.017} & 0.402 & \cellcolor{ebCoral!35}\textbf{0.000} & 0.365 & 0.486 \\
{\glm} & \cellcolor{ebCoral!35}\textbf{0.653}\,{\color{black!60}[0.600, 0.707]} & \cellcolor{ebCoral!35}\textbf{0.896} & 1.000 & \cellcolor{ebCoral!35}\textbf{0.017} & \cellcolor{ebCoral!35}\textbf{0.264} & 0.017 & 0.820 & 3.251 \\
{\qwenflash} & 0.470\,{\color{black!60}[0.413, 0.527]} & 0.848 & 1.000 & 0.050 & 0.428 & \cellcolor{ebCoral!35}\textbf{0.000} & 1.299 & 2.073 \\
{\qwenjson{} (ref.)} & 0.277\,{\color{black!60}[0.227, 0.330]} & 0.777 & 1.000 & 0.053 & 0.499 & 0.018 & 0.762 & 0.789 \\
\rowcolor{ebAmber!45}\multicolumn{9}{l}{\textbf{\textit{Key--value}}} \\
\rowcolor{ebCoral!10}{\jev} & 0.930\,{\color{black!60}[0.900, 0.957]} & 0.982 & 1.000 & 0.000 & 0.028 & 0.012 & 0.278 & 0.403 \\
{\semif} & 0.613\,{\color{black!60}[0.560, 0.670]} & 0.889 & 1.000 & 0.000 & 0.082 & 0.059 & 0.121 & 0.122 \\
{\laya} & 0.110\,{\color{black!60}[0.077, 0.147]} & 0.610 & 1.000 & 0.033 & 0.675 & 0.040 & \cellcolor{ebCoral!35}\textbf{0.055} & \cellcolor{ebCoral!35}\textbf{0.055} \\
{\deepseek} & \cellcolor{ebCoral!35}\textbf{0.987}\,{\color{black!60}[0.973, 0.997]} & \cellcolor{ebCoral!35}\textbf{0.997} & 1.000 & 0.000 & 0.005 & \cellcolor{ebCoral!35}\textbf{0.003} & 0.401 & 0.667 \\
{\glm} & 0.973\,{\color{black!60}[0.953, 0.990]} & 0.993 & 1.000 & 0.000 & \cellcolor{ebCoral!35}\textbf{0.002} & 0.005 & 0.907 & 3.082 \\
{\qwenflash} & 0.950\,{\color{black!60}[0.923, 0.973]} & 0.988 & 1.000 & 0.000 & 0.021 & \cellcolor{ebCoral!35}\textbf{0.003} & 1.278 & 2.125 \\
{\qwenjson{} (ref.)} & 0.610\,{\color{black!60}[0.553, 0.667]} & 0.887 & 1.000 & 0.000 & 0.179 & 0.023 & 0.763 & 0.809 \\
\rowcolor{ebAmber!45}\multicolumn{9}{l}{\textbf{\textit{Negation-heavy}}} \\
\rowcolor{ebCoral!10}{\jev} & 0.923\,{\color{black!60}[0.890, 0.950]} & 0.981 & 1.000 & 0.000 & 0.040 & 0.008 & 0.286 & 0.437 \\
{\semif} & 0.520\,{\color{black!60}[0.463, 0.577]} & 0.860 & 1.000 & 0.003 & 0.256 & 0.062 & 0.121 & 0.122 \\
{\laya} & 0.090\,{\color{black!60}[0.060, 0.123]} & 0.634 & 1.000 & 0.083 & 0.788 & 0.008 & \cellcolor{ebCoral!35}\textbf{0.055} & \cellcolor{ebCoral!35}\textbf{0.056} \\
{\deepseek} & \cellcolor{ebCoral!35}\textbf{0.973}\,{\color{black!60}[0.953, 0.990]} & \cellcolor{ebCoral!35}\textbf{0.993} & 1.000 & 0.000 & \cellcolor{ebCoral!35}\textbf{0.005} & 0.008 & 0.377 & 0.546 \\
{\glm} & 0.910\,{\color{black!60}[0.877, 0.940]} & 0.950 & 0.963 & 0.000 & \cellcolor{ebCoral!35}\textbf{0.005} & 0.054 & 0.809 & 3.969 \\
{\qwenflash} & 0.957\,{\color{black!60}[0.933, 0.977]} & 0.989 & 1.000 & 0.000 & 0.028 & \cellcolor{ebCoral!35}\textbf{0.001} & 1.522 & 2.716 \\
{\qwenjson{} (ref.)} & 0.433\,{\color{black!60}[0.377, 0.490]} & 0.823 & 1.000 & 0.000 & 0.322 & 0.036 & 0.762 & 0.811 \\
\rowcolor{ebAmber!45}\multicolumn{9}{l}{\textbf{\textit{Noisy}}} \\
\rowcolor{ebCoral!10}{\jev} & 0.930\,{\color{black!60}[0.900, 0.957]} & 0.983 & 1.000 & 0.000 & 0.045 & 0.003 & 0.285 & 0.399 \\
{\semif} & 0.557\,{\color{black!60}[0.500, 0.613]} & 0.877 & 1.000 & 0.000 & 0.186 & 0.053 & 0.121 & 0.122 \\
{\laya} & 0.097\,{\color{black!60}[0.063, 0.133]} & 0.645 & 1.000 & 0.127 & 0.739 & 0.014 & \cellcolor{ebCoral!35}\textbf{0.055} & \cellcolor{ebCoral!35}\textbf{0.056} \\
{\deepseek} & \cellcolor{ebCoral!35}\textbf{0.970}\,{\color{black!60}[0.950, 0.987]} & \cellcolor{ebCoral!35}\textbf{0.993} & 1.000 & 0.000 & 0.019 & 0.001 & 0.368 & 0.604 \\
{\glm} & 0.933\,{\color{black!60}[0.907, 0.960]} & 0.963 & 0.973 & 0.007 & \cellcolor{ebCoral!35}\textbf{0.014} & 0.032 & 0.671 & 2.728 \\
{\qwenflash} & 0.947\,{\color{black!60}[0.920, 0.970]} & 0.987 & 1.000 & 0.003 & 0.035 & \cellcolor{ebCoral!35}\textbf{0.000} & 1.535 & 2.750 \\
{\qwenjson{} (ref.)} & 0.463\,{\color{black!60}[0.410, 0.520]} & 0.838 & 1.000 & 0.000 & 0.292 & 0.035 & 0.763 & 0.811 \\
\rowcolor{ebAmber!45}\multicolumn{9}{l}{\textbf{\textit{Self-revising}}} \\
\rowcolor{ebCoral!10}{\jev} & 0.947\,{\color{black!60}[0.920, 0.970]} & 0.987 & 1.000 & 0.000 & 0.019 & 0.006 & 0.277 & 0.394 \\
{\semif} & 0.527\,{\color{black!60}[0.470, 0.583]} & 0.861 & 1.000 & 0.007 & 0.198 & 0.034 & 0.121 & 0.122 \\
{\laya} & 0.080\,{\color{black!60}[0.050, 0.113]} & 0.593 & 1.000 & 0.123 & 0.720 & 0.012 & \cellcolor{ebCoral!35}\textbf{0.055} & \cellcolor{ebCoral!35}\textbf{0.056} \\
{\deepseek} & \cellcolor{ebCoral!35}\textbf{0.990}\,{\color{black!60}[0.977, 1.000]} & \cellcolor{ebCoral!35}\textbf{0.998} & 1.000 & 0.000 & 0.007 & \cellcolor{ebCoral!35}\textbf{0.000} & 0.398 & 0.637 \\
{\glm} & 0.967\,{\color{black!60}[0.947, 0.983]} & 0.992 & 1.000 & 0.000 & \cellcolor{ebCoral!35}\textbf{0.000} & 0.008 & 0.871 & 2.816 \\
{\qwenflash} & 0.987\,{\color{black!60}[0.973, 0.997]} & 0.997 & 1.000 & 0.000 & 0.002 & \cellcolor{ebCoral!35}\textbf{0.000} & 1.456 & 2.855 \\
{\qwenjson{} (ref.)} & 0.510\,{\color{black!60}[0.453, 0.567]} & 0.850 & 1.000 & 0.000 & 0.242 & 0.023 & 0.765 & 0.983 \\
\bottomrule
\end{tabular}}
\def\ebh#1#2{#1 #2}\sbox0{\ebtab}
\ifdim\wd0>\dimexpr\linewidth-1pt\relax\def\ebh#1#2{\shortstack{#1\\{}#2}}\sbox0{\ebtab}\fi
\setlength{\tabcolsep}{\dimexpr\tabcolsep+(\linewidth-\wd0-1pt)/18\relax}
\ifdim\tabcolsep<1.5pt\setlength{\tabcolsep}{1.5pt}\fi
\typeout{EBTAB tab_rq2 natural=\the\wd0\space line=\the\linewidth\space sep=\the\tabcolsep}
\ebtab
\end{table*}

\begin{table*}[!t]
\centering
\caption{RQ2 paired contrasts against \jev per wording condition: differences in EM and in the unsafe-locality rate with 95\% confidence intervals, exact McNemar $p$, and the Holm-adjusted $p$ of H3.}
\label{tab:rq2_contrasts}
\scriptsize
\setlength{\tabcolsep}{6pt}
\setlength{\aboverulesep}{0pt}\setlength{\belowrulesep}{0pt}\setlength{\extrarowheight}{0pt}
\renewcommand{\arraystretch}{0.94}
\def\ebtab{%
\begin{tabular}{lrrrrrl}
\toprule
\rowcolor{ebBlue}Contrast & \ebh{EM diff.}{[95\% CI]} & \ebh{EM}{McNemar $p$} & \ebh{Unsafe diff.}{[95\% CI]} & \ebh{Unsafe}{McNemar $p$} & \ebh{Unsafe Holm}{$p$ (H3)} & Verdict \\
\midrule
\rowcolor{ebAmber!45}\multicolumn{7}{l}{\textbf{\textit{Clean}}} \\
{\semif} $-$ {\jev} & $-$0.310\,{\color{black!60}[$-$0.367, $-$0.253]} & 8.4e-23 & 0.003\,{\color{black!60}[0.000, 0.010]} & 1 & 1 & not resolved \\
{\laya} $-$ {\jev} & $-$0.840\,{\color{black!60}[$-$0.880, $-$0.797]} & 2.8e-76 & 0.130\,{\color{black!60}[0.093, 0.170]} & 3.6e-12 & 1.5e-10 & \cellcolor{ebCoral!35}\textbf{resolved} \\
{\deepseek} $-$ {\jev} & 0.037\,{\color{black!60}[0.013, 0.063]} & 0.0074 & 0.000\,{\color{black!60}[0.000, 0.000]} & 1 & 1 & not resolved \\
{\glm} $-$ {\jev} & 0.043\,{\color{black!60}[0.017, 0.070]} & 0.0023 & 0.000\,{\color{black!60}[0.000, 0.000]} & 1 & 1 & not resolved \\
{\qwenflash} $-$ {\jev} & 0.010\,{\color{black!60}[$-$0.017, 0.037]} & 0.63 & 0.000\,{\color{black!60}[0.000, 0.000]} & 1 & 1 & not resolved \\
\rowcolor{ebAmber!45}\multicolumn{7}{l}{\textbf{\textit{Code-switched}}} \\
{\semif} $-$ {\jev} & $-$0.397\,{\color{black!60}[$-$0.457, $-$0.337]} & 2.4e-29 & 0.003\,{\color{black!60}[0.000, 0.010]} & 1 & 1 & not resolved \\
{\laya} $-$ {\jev} & $-$0.853\,{\color{black!60}[$-$0.893, $-$0.810]} & 1.7e-77 & 0.123\,{\color{black!60}[0.087, 0.160]} & 1.5e-11 & 5.5e-10 & \cellcolor{ebCoral!35}\textbf{resolved} \\
{\deepseek} $-$ {\jev} & 0.063\,{\color{black!60}[0.033, 0.097]} & 1.6e-4 & 0.000\,{\color{black!60}[0.000, 0.000]} & 1 & 1 & not resolved \\
{\glm} $-$ {\jev} & 0.043\,{\color{black!60}[0.007, 0.083]} & 0.035 & 0.003\,{\color{black!60}[0.000, 0.010]} & 1 & 1 & not resolved \\
{\qwenflash} $-$ {\jev} & 0.043\,{\color{black!60}[0.010, 0.077]} & 0.019 & 0.000\,{\color{black!60}[0.000, 0.000]} & 1 & 1 & not resolved \\
\rowcolor{ebAmber!45}\multicolumn{7}{l}{\textbf{\textit{Colloquial}}} \\
{\semif} $-$ {\jev} & $-$0.437\,{\color{black!60}[$-$0.493, $-$0.380]} & 2.5e-38 & 0.007\,{\color{black!60}[0.000, 0.017]} & 0.5 & 1 & not resolved \\
{\laya} $-$ {\jev} & $-$0.840\,{\color{black!60}[$-$0.880, $-$0.797]} & 1.8e-74 & 0.073\,{\color{black!60}[0.047, 0.103]} & 4.8e-7 & 1.7e-5 & \cellcolor{ebCoral!35}\textbf{resolved} \\
{\deepseek} $-$ {\jev} & 0.053\,{\color{black!60}[0.027, 0.083]} & 1.4e-4 & 0.000\,{\color{black!60}[0.000, 0.000]} & 1 & 1 & not resolved \\
{\glm} $-$ {\jev} & 0.040\,{\color{black!60}[0.010, 0.073]} & 0.023 & 0.000\,{\color{black!60}[0.000, 0.000]} & 1 & 1 & not resolved \\
{\qwenflash} $-$ {\jev} & 0.020\,{\color{black!60}[$-$0.013, 0.057]} & 0.34 & 0.000\,{\color{black!60}[0.000, 0.000]} & 1 & 1 & not resolved \\
\rowcolor{ebAmber!45}\multicolumn{7}{l}{\textbf{\textit{Default-baiting}}} \\
{\semif} $-$ {\jev} & $-$0.243\,{\color{black!60}[$-$0.303, $-$0.180]} & 3.5e-13 & $-$0.023\,{\color{black!60}[$-$0.047, 0.000]} & 0.092 & 1 & not resolved \\
{\laya} $-$ {\jev} & $-$0.547\,{\color{black!60}[$-$0.607, $-$0.487]} & 1.2e-44 & 0.090\,{\color{black!60}[0.050, 0.133]} & 4.2e-5 & 0.0014 & \cellcolor{ebCoral!35}\textbf{resolved} \\
{\deepseek} $-$ {\jev} & $-$0.080\,{\color{black!60}[$-$0.137, $-$0.023]} & 0.0071 & $-$0.027\,{\color{black!60}[$-$0.050, $-$0.007]} & 0.039 & 1 & not resolved \\
{\glm} $-$ {\jev} & 0.043\,{\color{black!60}[$-$0.007, 0.093]} & 0.12 & $-$0.027\,{\color{black!60}[$-$0.050, $-$0.007]} & 0.039 & 1 & not resolved \\
{\qwenflash} $-$ {\jev} & $-$0.140\,{\color{black!60}[$-$0.197, $-$0.083]} & 3.7e-6 & 0.007\,{\color{black!60}[$-$0.017, 0.030]} & 0.79 & 1 & not resolved \\
\rowcolor{ebAmber!45}\multicolumn{7}{l}{\textbf{\textit{Key--value}}} \\
{\semif} $-$ {\jev} & $-$0.317\,{\color{black!60}[$-$0.373, $-$0.260]} & 5.0e-24 & 0.000\,{\color{black!60}[0.000, 0.000]} & 1 & 1 & not resolved \\
{\laya} $-$ {\jev} & $-$0.820\,{\color{black!60}[$-$0.863, $-$0.773]} & 1.1e-72 & 0.033\,{\color{black!60}[0.013, 0.057]} & 0.002 & 0.064 & not resolved \\
{\deepseek} $-$ {\jev} & 0.057\,{\color{black!60}[0.030, 0.087]} & 7.6e-5 & 0.000\,{\color{black!60}[0.000, 0.000]} & 1 & 1 & not resolved \\
{\glm} $-$ {\jev} & 0.043\,{\color{black!60}[0.013, 0.073]} & 0.011 & 0.000\,{\color{black!60}[0.000, 0.000]} & 1 & 1 & not resolved \\
{\qwenflash} $-$ {\jev} & 0.020\,{\color{black!60}[$-$0.013, 0.053]} & 0.31 & 0.000\,{\color{black!60}[0.000, 0.000]} & 1 & 1 & not resolved \\
\rowcolor{ebAmber!45}\multicolumn{7}{l}{\textbf{\textit{Negation-heavy}}} \\
{\semif} $-$ {\jev} & $-$0.403\,{\color{black!60}[$-$0.463, $-$0.347]} & 4.0e-33 & 0.003\,{\color{black!60}[0.000, 0.010]} & 1 & 1 & not resolved \\
{\laya} $-$ {\jev} & $-$0.833\,{\color{black!60}[$-$0.873, $-$0.793]} & 1.1e-75 & 0.083\,{\color{black!60}[0.053, 0.117]} & 6.0e-8 & 2.1e-6 & \cellcolor{ebCoral!35}\textbf{resolved} \\
{\deepseek} $-$ {\jev} & 0.050\,{\color{black!60}[0.017, 0.083]} & 0.0059 & 0.000\,{\color{black!60}[0.000, 0.000]} & 1 & 1 & not resolved \\
{\glm} $-$ {\jev} & $-$0.013\,{\color{black!60}[$-$0.057, 0.030]} & 0.64 & 0.000\,{\color{black!60}[0.000, 0.000]} & 1 & 1 & not resolved \\
{\qwenflash} $-$ {\jev} & 0.033\,{\color{black!60}[$-$0.003, 0.070]} & 0.099 & 0.000\,{\color{black!60}[0.000, 0.000]} & 1 & 1 & not resolved \\
\rowcolor{ebAmber!45}\multicolumn{7}{l}{\textbf{\textit{Noisy}}} \\
{\semif} $-$ {\jev} & $-$0.373\,{\color{black!60}[$-$0.433, $-$0.317]} & 1.3e-29 & 0.000\,{\color{black!60}[0.000, 0.000]} & 1 & 1 & not resolved \\
{\laya} $-$ {\jev} & $-$0.833\,{\color{black!60}[$-$0.873, $-$0.790]} & 1.1e-75 & 0.127\,{\color{black!60}[0.090, 0.163]} & 7.3e-12 & 2.8e-10 & \cellcolor{ebCoral!35}\textbf{resolved} \\
{\deepseek} $-$ {\jev} & 0.040\,{\color{black!60}[0.010, 0.073]} & 0.023 & 0.000\,{\color{black!60}[0.000, 0.000]} & 1 & 1 & not resolved \\
{\glm} $-$ {\jev} & 0.003\,{\color{black!60}[$-$0.033, 0.040]} & 1 & 0.007\,{\color{black!60}[0.000, 0.017]} & 0.5 & 1 & not resolved \\
{\qwenflash} $-$ {\jev} & 0.017\,{\color{black!60}[$-$0.017, 0.050]} & 0.44 & 0.003\,{\color{black!60}[0.000, 0.010]} & 1 & 1 & not resolved \\
\rowcolor{ebAmber!45}\multicolumn{7}{l}{\textbf{\textit{Self-revising}}} \\
{\semif} $-$ {\jev} & $-$0.420\,{\color{black!60}[$-$0.480, $-$0.363]} & 1.4e-34 & 0.007\,{\color{black!60}[0.000, 0.017]} & 0.5 & 1 & not resolved \\
{\laya} $-$ {\jev} & $-$0.867\,{\color{black!60}[$-$0.907, $-$0.827]} & 7.1e-77 & 0.123\,{\color{black!60}[0.087, 0.163]} & 1.5e-11 & 5.5e-10 & \cellcolor{ebCoral!35}\textbf{resolved} \\
{\deepseek} $-$ {\jev} & 0.043\,{\color{black!60}[0.017, 0.070]} & 0.0023 & 0.000\,{\color{black!60}[0.000, 0.000]} & 1 & 1 & not resolved \\
{\glm} $-$ {\jev} & 0.020\,{\color{black!60}[$-$0.007, 0.050]} & 0.24 & 0.000\,{\color{black!60}[0.000, 0.000]} & 1 & 1 & not resolved \\
{\qwenflash} $-$ {\jev} & 0.040\,{\color{black!60}[0.013, 0.067]} & 0.0075 & 0.000\,{\color{black!60}[0.000, 0.000]} & 1 & 1 & not resolved \\
\bottomrule
\end{tabular}}
\def\ebh#1#2{#1 #2}\sbox0{\ebtab}
\ifdim\wd0>\dimexpr\linewidth-1pt\relax\def\ebh#1#2{\shortstack{#1\\{}#2}}\sbox0{\ebtab}\fi
\setlength{\tabcolsep}{\dimexpr\tabcolsep+(\linewidth-\wd0-1pt)/14\relax}
\ifdim\tabcolsep<1.5pt\setlength{\tabcolsep}{1.5pt}\fi
\typeout{EBTAB tab_rq2_contrasts natural=\the\wd0\space line=\the\linewidth\space sep=\the\tabcolsep}
\ebtab
\end{table*}

Most wording changes cost little accuracy. Leaving out the default-baiting
condition, \jev's exact match stays at 0.920--0.950 across the other seven
conditions and \deepseek's at 0.970--0.990 (Fig.~\ref{fig:rq2} and
Table~\ref{tab:rq2}). Default-baiting wording, which hints at values that the
request never states, lowers every interpreter. \glm keeps 0.653, \jev 0.610,
\deepseek 0.530, and \qwenflash 0.470. The failures are spurious
specifications, values that the reference leaves unspecified, at rates of
0.264--0.428 for the four hosted interpreters.

Unsafe decisions differ across the six interpreters in all eight conditions
(Cochran's $Q$, H3). The difference comes from \laya, whose unsafe rate is
0.033--0.133 and exceeds \jev's in seven of eight paired contrasts
(Table~\ref{tab:rq2_contrasts}). \jev
makes no unsafe decision in seven conditions and 4.3\% under default-baiting,
and the hosted LLMs stay at or below 5.0\%. None of the paired contrasts
between \jev and an LLM or \semif resolves.

\takeaway{Input quality}{Because all four hosted interpreters keep unsafe
placement at or below 5\% under every wording, wording robustness does not
separate them. Default-baiting is the shared weak point, and there \jev ranks
second of the six.}

\subsection{RQ3: Task Difficulty}
\label{sec:rq3}
\begin{figure*}[!t]
\centering
\includegraphics[width=7.00in]{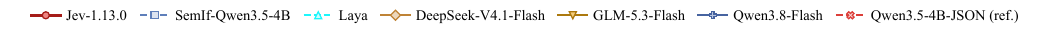}\\[-2pt]
\subfloat[EM vs.\ $F$ ($D$ = medium)]{\includegraphics[width=1.70in]{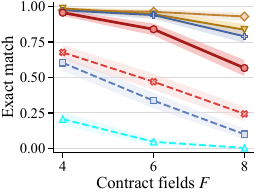}\label{fig:rq3_a}}\hfill
\subfloat[p50 latency vs.\ $F$ ($D$ = medium)]{\includegraphics[width=1.70in]{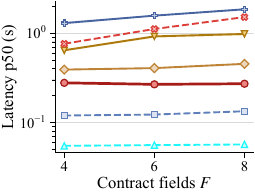}\label{fig:rq3_b}}\hfill
\subfloat[Output tokens ($D$ = medium)]{\includegraphics[width=1.70in]{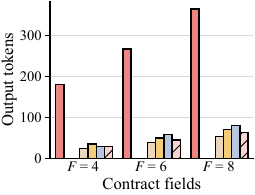}\label{fig:rq3_c}}\hfill
\subfloat[EM vs.\ $D$ ($F$ = 8)]{\includegraphics[width=1.70in]{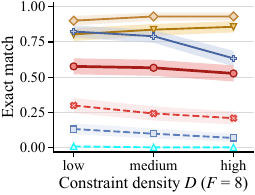}\label{fig:rq3_d}}
\caption{RQ3, task difficulty. (a) EM, (b) median decision latency, and (c) median output tokens as the contract grows from four to eight fields at medium constraint density; (d) EM across constraint density with eight fields.}
\label{fig:rq3}
\end{figure*}

\begin{table*}[!t]
\centering
\caption{RQ3, task difficulty. EM and median decision latency for each contract size $F$ and constraint density $D$, and median output tokens at medium density.}
\label{tab:rq3}
\scriptsize
\setlength{\tabcolsep}{7pt}
\setlength{\aboverulesep}{0pt}\setlength{\belowrulesep}{0pt}\setlength{\extrarowheight}{0pt}
\renewcommand{\arraystretch}{0.94}
\def\ebtab{%
\begin{tabular}{lrrrrrrr}
\toprule
\rowcolor{ebBlue} & \multicolumn{3}{c}{EM $\uparrow$ by constraint density $D$} & \multicolumn{3}{c}{p50 (s) $\downarrow$ by constraint density $D$} & Out.\ tokens \\
\cmidrule(lr){2-4}\cmidrule(lr){5-7}
\rowcolor{ebBlue}Model & low & medium & high & low & medium & high & $D$ = medium \\
\midrule
\rowcolor{ebAmber!45}\multicolumn{8}{l}{\textbf{\textit{$F=4$ contract fields}}} \\
\rowcolor{ebCoral!10}{\jev} & 0.923 & 0.957 & 0.967 & 0.279 & 0.278 & 0.268 & 180 \\
{\semif} & 0.543 & 0.603 & 0.650 & 0.120 & 0.120 & 0.122 & 0 \\
{\laya} & 0.253 & 0.207 & 0.157 & \cellcolor{ebCoral!35}\textbf{0.055} & \cellcolor{ebCoral!35}\textbf{0.055} & \cellcolor{ebCoral!35}\textbf{0.056} & 0 \\
{\deepseek} & \cellcolor{ebCoral!35}\textbf{0.983} & 0.973 & 0.973 & 0.417 & 0.392 & 0.372 & 25 \\
{\glm} & 0.897 & \cellcolor{ebCoral!35}\textbf{0.987} & \cellcolor{ebCoral!35}\textbf{0.997} & 0.916 & 0.645 & 0.645 & 35 \\
{\qwenflash} & \cellcolor{ebCoral!35}\textbf{0.983} & 0.973 & 0.987 & 1.275 & 1.301 & 1.325 & 30 \\
{\qwenjson{} (ref.)} & 0.530 & 0.677 & 0.677 & 0.766 & 0.764 & 0.767 & 30 \\
\rowcolor{ebAmber!45}\multicolumn{8}{l}{\textbf{\textit{$F=6$ contract fields}}} \\
\rowcolor{ebCoral!10}{\jev} & 0.787 & 0.840 & 0.780 & 0.280 & 0.268 & 0.280 & 267 \\
{\semif} & 0.400 & 0.337 & 0.360 & 0.124 & 0.123 & 0.124 & 0 \\
{\laya} & 0.077 & 0.047 & 0.040 & \cellcolor{ebCoral!35}\textbf{0.056} & \cellcolor{ebCoral!35}\textbf{0.056} & \cellcolor{ebCoral!35}\textbf{0.056} & 0 \\
{\deepseek} & 0.957 & \cellcolor{ebCoral!35}\textbf{0.963} & \cellcolor{ebCoral!35}\textbf{0.987} & 0.415 & 0.407 & 0.404 & 39 \\
{\glm} & \cellcolor{ebCoral!35}\textbf{0.963} & 0.947 & 0.913 & 0.704 & 0.923 & 0.895 & 50 \\
{\qwenflash} & 0.900 & 0.940 & 0.953 & 1.792 & 1.577 & 1.813 & 59 \\
{\qwenjson{} (ref.)} & 0.483 & 0.470 & 0.523 & 1.105 & 1.118 & 1.112 & 45 \\
\rowcolor{ebAmber!45}\multicolumn{8}{l}{\textbf{\textit{$F=8$ contract fields}}} \\
\rowcolor{ebCoral!10}{\jev} & 0.577 & 0.567 & 0.527 & 0.272 & 0.272 & 0.264 & 364 \\
{\semif} & 0.133 & 0.100 & 0.070 & 0.133 & 0.134 & 0.135 & 0 \\
{\laya} & 0.010 & 0.003 & 0.003 & \cellcolor{ebCoral!35}\textbf{0.057} & \cellcolor{ebCoral!35}\textbf{0.057} & \cellcolor{ebCoral!35}\textbf{0.057} & 0 \\
{\deepseek} & \cellcolor{ebCoral!35}\textbf{0.900} & \cellcolor{ebCoral!35}\textbf{0.930} & \cellcolor{ebCoral!35}\textbf{0.930} & 0.460 & 0.455 & 0.483 & 53 \\
{\glm} & 0.803 & 0.837 & 0.857 & 0.930 & 0.983 & 1.272 & 71 \\
{\qwenflash} & 0.823 & 0.790 & 0.633 & 2.092 & 1.859 & 1.965 & 81 \\
{\qwenjson{} (ref.)} & 0.300 & 0.243 & 0.210 & 1.507 & 1.519 & 1.547 & 63 \\
\bottomrule
\end{tabular}}
\def\ebh#1#2{#1 #2}\sbox0{\ebtab}
\ifdim\wd0>\dimexpr\linewidth-1pt\relax\def\ebh#1#2{\shortstack{#1\\{}#2}}\sbox0{\ebtab}\fi
\setlength{\tabcolsep}{\dimexpr\tabcolsep+(\linewidth-\wd0-1pt)/16\relax}
\ifdim\tabcolsep<1.5pt\setlength{\tabcolsep}{1.5pt}\fi
\typeout{EBTAB tab_rq3 natural=\the\wd0\space line=\the\linewidth\space sep=\the\tabcolsep}
\ebtab
\end{table*}

Contract size is where the decision model pays. With four fields, \jev reaches
an exact match of 0.923--0.967, close to the LLMs. With six fields it falls to
0.780--0.840, and with eight to 0.527--0.577, while \deepseek keeps
0.900--0.930 (Fig.~\ref{fig:rq3} and Table~\ref{tab:rq3}). Constraint density
moves exact match much less than the number of fields.

The LLMs' median output grows by 28, 36, and 43 tokens from four to eight
fields, and all three contrasts resolve (H4). Their latency grows with it and
spans 0.372--0.483\,s over the nine cells for \deepseek and 1.275--2.092\,s for
\qwenflash. \jev's latency stays at 0.264--0.280\,s in every cell. Its
growth from four to eight fields is 0.83, 0.64, and 0.65 times that of the
three LLMs, and 0.56 for \semif against \qwenjson. These four contrasts all resolve.

\takeaway{Task difficulty}{Each added field costs an LLM output tokens and time
but leaves \jev's time unchanged. At eight fields the trade reverses. \deepseek
is then 32.3--40.3 points more accurate for about 0.2\,s of additional latency.}

\subsection{RQ4: Dynamic Service Catalog}
\label{sec:rq4}
\begin{figure*}[!t]
\centering
\includegraphics[width=7.00in]{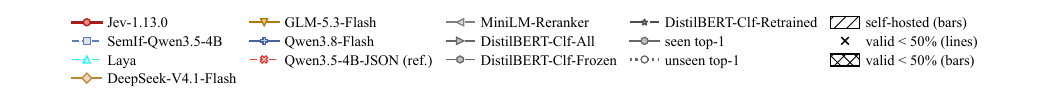}\\[-2pt]
\subfloat[Seen and unseen top-1 vs.\ $K$]{\includegraphics[width=1.70in]{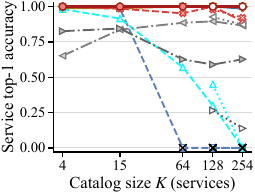}\label{fig:rq4_a}}\hfill
\subfloat[p50 latency vs.\ $K$]{\includegraphics[width=1.70in]{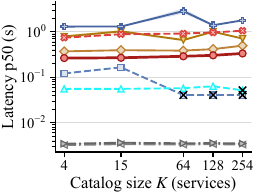}\label{fig:rq4_b}}\hfill
\subfloat[Top-1 under churn]{\includegraphics[width=1.70in]{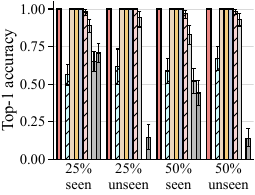}\label{fig:rq4_c}}\hfill
\subfloat[Unsupported F1 vs.\ $K$]{\includegraphics[width=1.70in]{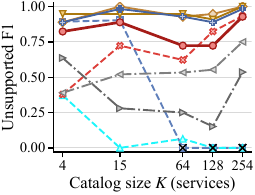}\label{fig:rq4_d}}
\caption{RQ4, dynamic service catalog. (a) Top-1 service accuracy on seen (solid) and unseen (dotted) services and (b) median decision latency as the catalog grows from 4 to 254 services; (c) top-1 accuracy when 25\% or 50\% of a 64-service catalog is replaced; (d) F1 score for detecting unsupported requests. Crosses mark cells with fewer than half valid outputs.}
\label{fig:rq4}
\end{figure*}

\begin{table*}[!t]
\centering
\caption{RQ4a, catalog size. Service accuracy on seen and unseen services, unsupported-request detection, admission errors, EM, validity, and latency for catalogs of 4 to 254 services, with the trained classifier and the sentence-embedding reranker as references.}
\label{tab:rq4a}
\scriptsize
\setlength{\tabcolsep}{5pt}
\setlength{\aboverulesep}{0pt}\setlength{\belowrulesep}{0pt}\setlength{\extrarowheight}{0pt}
\renewcommand{\arraystretch}{0.94}
\providecommand{\ebflagnote}{$^\dagger$Valid output rate below 0.5; the row is printed but excluded from the best-value marking.}
\def\ebtab{%
\begin{tabular}{lrrrrrrrrr}
\toprule
\rowcolor{ebBlue}Model & \ebh{Seen}{top-1 $\uparrow$} & \ebh{Unseen}{top-1 $\uparrow$} & \ebh{Unsupported}{F1 $\uparrow$} & \ebh{False}{admission $\downarrow$} & \ebh{False}{blocking $\downarrow$} & \ebh{EM}{$\uparrow$} & \ebh{Valid}{$\uparrow$} & \ebh{p50}{(s) $\downarrow$} & \ebh{p95}{(s) $\downarrow$} \\
\midrule
\rowcolor{ebAmber!45}\multicolumn{10}{l}{\textbf{\textit{$K=4$ services}}} \\
\rowcolor{ebCoral!10}{\jev} & \cellcolor{ebCoral!35}\textbf{1.000} & -- & 0.824 & 0.300 & 0.000 & 0.920 & 1.000 & 0.266 & 0.351 \\
{\semif} & 0.996 & -- & 0.893 & 0.167 & 0.004 & 0.620 & 1.000 & 0.122 & 0.213 \\
{\laya} & 0.981 & -- & 0.372 & 0.733 & 0.019 & 0.207 & 1.000 & 0.056 & 0.151 \\
{\deepseek} & \cellcolor{ebCoral!35}\textbf{1.000} & -- & 0.889 & 0.200 & 0.000 & \cellcolor{ebCoral!35}\textbf{0.957} & 1.000 & 0.372 & 0.707 \\
{\glm} & 0.989 & -- & \cellcolor{ebCoral!35}\textbf{0.947} & 0.100 & 0.000 & 0.947 & 0.987 & 0.790 & 2.944 \\
{\qwenflash} & 0.996 & -- & 0.889 & 0.200 & 0.000 & 0.953 & 0.997 & 1.308 & 2.413 \\
{\qwenjson{} (ref.)} & \cellcolor{ebCoral!35}\textbf{1.000} & -- & 0.378 & 0.767 & 0.000 & 0.467 & 1.000 & 0.744 & 0.879 \\
{MiniLM-Reranker} & 0.652 & -- & 0.390 & \cellcolor{ebCoral!35}\textbf{0.000} & 0.348 & -- & 1.000 & \cellcolor{ebCoral!35}\textbf{0.003} & \cellcolor{ebCoral!35}\textbf{0.005} \\
{DistilBERT-Clf-All} & 0.826 & -- & 0.636 & 0.533 & 0.000 & -- & 1.000 & \cellcolor{ebCoral!35}\textbf{0.003} & 0.006 \\
\rowcolor{ebAmber!45}\multicolumn{10}{l}{\textbf{\textit{$K=15$ services}}} \\
\rowcolor{ebCoral!10}{\jev} & \cellcolor{ebCoral!35}\textbf{1.000} & -- & 0.889 & 0.200 & 0.000 & 0.907 & 1.000 & 0.269 & 0.374 \\
{\semif} & 0.985 & -- & 0.903 & 0.067 & 0.015 & 0.643 & 1.000 & 0.166 & 0.246 \\
{\laya} & 0.915 & -- & 0.000 & 1.000 & 0.019 & 0.183 & 1.000 & 0.056 & 0.057 \\
{\deepseek} & \cellcolor{ebCoral!35}\textbf{1.000} & -- & \cellcolor{ebCoral!35}\textbf{1.000} & \cellcolor{ebCoral!35}\textbf{0.000} & 0.000 & \cellcolor{ebCoral!35}\textbf{0.987} & 1.000 & 0.395 & 0.608 \\
{\glm} & 0.989 & -- & 0.947 & 0.100 & 0.000 & 0.967 & 0.987 & 1.027 & 4.191 \\
{\qwenflash} & 0.996 & -- & 0.983 & 0.033 & 0.000 & 0.950 & 0.997 & 1.310 & 2.259 \\
{\qwenjson{} (ref.)} & 0.985 & -- & 0.723 & 0.433 & 0.000 & 0.457 & 1.000 & 0.888 & 1.003 \\
{MiniLM-Reranker} & 0.837 & -- & 0.520 & 0.133 & 0.163 & -- & 1.000 & \cellcolor{ebCoral!35}\textbf{0.003} & \cellcolor{ebCoral!35}\textbf{0.006} \\
{DistilBERT-Clf-All} & 0.844 & -- & 0.280 & 0.767 & 0.048 & -- & 1.000 & 0.004 & 0.008 \\
\rowcolor{ebAmber!45}\multicolumn{10}{l}{\textbf{\textit{$K=64$ services}}} \\
\rowcolor{ebCoral!10}{\jev} & \cellcolor{ebCoral!35}\textbf{1.000} & -- & 0.723 & 0.433 & 0.000 & 0.867 & 1.000 & 0.290 & 0.378 \\
\cellcolor{ebPink}{\semif}$^\dagger$ & \cellcolor{ebPink}0.000 & -- & \cellcolor{ebPink}0.000 & \cellcolor{ebPink}1.000 & \cellcolor{ebPink}0.000 & \cellcolor{ebPink}0.000 & \cellcolor{ebPink}0.000 & \cellcolor{ebPink}0.041 & \cellcolor{ebPink}0.041 \\
{\laya} & 0.570 & -- & 0.065 & 0.967 & 0.000 & 0.107 & 1.000 & 0.058 & 0.059 \\
{\deepseek} & \cellcolor{ebCoral!35}\textbf{1.000} & -- & 0.929 & 0.133 & 0.000 & 0.953 & 1.000 & 0.387 & 0.570 \\
{\glm} & \cellcolor{ebCoral!35}\textbf{1.000} & -- & \cellcolor{ebCoral!35}\textbf{0.947} & \cellcolor{ebCoral!35}\textbf{0.100} & 0.000 & \cellcolor{ebCoral!35}\textbf{0.973} & 1.000 & 0.661 & 1.232 \\
{\qwenflash} & \cellcolor{ebCoral!35}\textbf{1.000} & -- & 0.929 & 0.133 & 0.000 & 0.963 & 1.000 & 2.861 & 7.568 \\
{\qwenjson{} (ref.)} & 0.952 & -- & 0.622 & 0.533 & 0.004 & 0.437 & 1.000 & 0.918 & 1.011 \\
{MiniLM-Reranker} & 0.885 & -- & 0.533 & 0.467 & 0.052 & -- & 1.000 & \cellcolor{ebCoral!35}\textbf{0.003} & \cellcolor{ebCoral!35}\textbf{0.005} \\
{DistilBERT-Clf-All} & 0.626 & -- & 0.250 & 0.733 & 0.096 & -- & 1.000 & 0.004 & 0.007 \\
\rowcolor{ebAmber!45}\multicolumn{10}{l}{\textbf{\textit{$K=128$ services}}} \\
\rowcolor{ebCoral!10}{\jev} & 1.000 & \cellcolor{ebCoral!35}\textbf{1.000} & 0.723 & 0.433 & 0.000 & 0.887 & 1.000 & 0.305 & 0.406 \\
\cellcolor{ebPink}{\semif}$^\dagger$ & \cellcolor{ebPink}0.000 & \cellcolor{ebPink}0.000 & \cellcolor{ebPink}0.000 & \cellcolor{ebPink}1.000 & \cellcolor{ebPink}0.000 & \cellcolor{ebPink}0.000 & \cellcolor{ebPink}0.000 & \cellcolor{ebPink}0.041 & \cellcolor{ebPink}0.041 \\
{\laya} & 0.299 & 0.449 & 0.000 & 1.000 & 0.007 & 0.083 & 1.000 & 0.064 & 0.172 \\
{\deepseek} & 1.000 & \cellcolor{ebCoral!35}\textbf{1.000} & \cellcolor{ebCoral!35}\textbf{0.947} & \cellcolor{ebCoral!35}\textbf{0.100} & 0.000 & \cellcolor{ebCoral!35}\textbf{0.967} & 1.000 & 0.417 & 0.999 \\
{\glm} & 1.000 & \cellcolor{ebCoral!35}\textbf{1.000} & 0.909 & 0.167 & 0.000 & 0.963 & 1.000 & 1.009 & 2.736 \\
{\qwenflash} & 1.000 & \cellcolor{ebCoral!35}\textbf{1.000} & 0.889 & 0.200 & 0.000 & 0.943 & 1.000 & 1.399 & 2.035 \\
{\qwenjson{} (ref.)} & 1.000 & 0.978 & 0.824 & 0.300 & 0.000 & 0.417 & 1.000 & 0.971 & 1.094 \\
{MiniLM-Reranker} & 0.896 & 0.926 & 0.553 & 0.567 & 0.015 & -- & 1.000 & \cellcolor{ebCoral!35}\textbf{0.003} & \cellcolor{ebCoral!35}\textbf{0.006} \\
{DistilBERT-Clf-All} & 0.590 & 0.265 & 0.152 & 0.833 & 0.115 & -- & 1.000 & 0.004 & 0.008 \\
\rowcolor{ebAmber!45}\multicolumn{10}{l}{\textbf{\textit{$K=254$ services}}} \\
\rowcolor{ebCoral!10}{\jev} & \cellcolor{ebCoral!35}\textbf{1.000} & \cellcolor{ebCoral!35}\textbf{1.000} & 0.929 & 0.133 & 0.000 & 0.937 & 1.000 & 0.335 & 0.432 \\
\cellcolor{ebPink}{\semif}$^\dagger$ & \cellcolor{ebPink}0.000 & \cellcolor{ebPink}0.000 & \cellcolor{ebPink}0.000 & \cellcolor{ebPink}1.000 & \cellcolor{ebPink}0.000 & \cellcolor{ebPink}0.000 & \cellcolor{ebPink}0.000 & \cellcolor{ebPink}0.041 & \cellcolor{ebPink}0.041 \\
\cellcolor{ebPink}{\laya}$^\dagger$ & \cellcolor{ebPink}0.000 & \cellcolor{ebPink}0.000 & \cellcolor{ebPink}0.000 & \cellcolor{ebPink}1.000 & \cellcolor{ebPink}0.000 & \cellcolor{ebPink}0.000 & \cellcolor{ebPink}0.000 & \cellcolor{ebPink}0.052 & \cellcolor{ebPink}0.052 \\
{\deepseek} & \cellcolor{ebCoral!35}\textbf{1.000} & 0.995 & \cellcolor{ebCoral!35}\textbf{1.000} & \cellcolor{ebCoral!35}\textbf{0.000} & 0.000 & 0.970 & 1.000 & 0.498 & 1.008 \\
{\glm} & \cellcolor{ebCoral!35}\textbf{1.000} & \cellcolor{ebCoral!35}\textbf{1.000} & \cellcolor{ebCoral!35}\textbf{1.000} & \cellcolor{ebCoral!35}\textbf{0.000} & 0.000 & \cellcolor{ebCoral!35}\textbf{0.980} & 1.000 & 0.711 & 1.876 \\
{\qwenflash} & 0.985 & \cellcolor{ebCoral!35}\textbf{1.000} & 0.983 & 0.033 & 0.000 & 0.963 & 1.000 & 1.795 & 3.791 \\
{\qwenjson{} (ref.)} & 0.896 & 0.921 & 0.933 & 0.067 & 0.007 & 0.493 & 1.000 & 1.059 & 1.179 \\
{MiniLM-Reranker} & 0.866 & 0.872 & 0.750 & 0.300 & 0.019 & -- & 1.000 & \cellcolor{ebCoral!35}\textbf{0.003} & \cellcolor{ebCoral!35}\textbf{0.004} \\
{DistilBERT-Clf-All} & 0.627 & 0.138 & 0.536 & 0.133 & 0.152 & -- & 1.000 & \cellcolor{ebCoral!35}\textbf{0.003} & \cellcolor{ebCoral!35}\textbf{0.004} \\
\bottomrule
\end{tabular}}
\def\ebh#1#2{#1 #2}\sbox0{\ebtab}
\ifdim\wd0>\dimexpr\linewidth-1pt\relax\def\ebh#1#2{\shortstack{#1\\{}#2}}\sbox0{\ebtab}\fi
\setlength{\tabcolsep}{\dimexpr\tabcolsep+(\linewidth-\wd0-1pt)/20\relax}
\ifdim\tabcolsep<1.5pt\setlength{\tabcolsep}{1.5pt}\fi
\typeout{EBTAB tab_rq4a natural=\the\wd0\space line=\the\linewidth\space sep=\the\tabcolsep}
\ebtab
\par\smallskip{\scriptsize\ebflagnote\par}
\end{table*}

\begin{table}[!t]
\centering
\caption{RQ4b, catalog churn at 64 services. Seen and unseen top-1 accuracy, their gap (H5), and the adaptation cost of the retrained classifier.}
\label{tab:rq4b}
\scriptsize
\setlength{\tabcolsep}{1.5pt}
\setlength{\aboverulesep}{0pt}\setlength{\belowrulesep}{0pt}\setlength{\extrarowheight}{0pt}
\renewcommand{\arraystretch}{0.94}
\providecommand{\ebflagnote}{$^\dagger$Valid output rate below 0.5; the row is printed but excluded from the best-value marking.}
\def\ebtab{%
\begin{tabular}{lrrrrr}
\toprule
\rowcolor{ebBlue}Model & \ebh{Seen}{$\uparrow$} & \ebh{Unseen}{$\uparrow$} & \ebh{Gap}{[95\% CI] $\to0$} & \ebh{Adapt.}{ex.} & \ebh{Adapt.}{(s)} \\
\midrule
\rowcolor{ebAmber!45}\multicolumn{6}{l}{\textbf{\textit{25\% churn}}} \\
\rowcolor{ebCoral!10}{\jev} & \cellcolor{ebCoral!35}\textbf{1.000} & 1.000 & \cellcolor{ebCoral!35}\textbf{0.000}\,{\color{black!60}[0.000, 0.000]} & -- & -- \\
\cellcolor{ebPink}{\semif}$^\dagger$ & \cellcolor{ebPink}0.000 & \cellcolor{ebPink}0.000 & \cellcolor{ebPink}0.000\,{\color{black!60}[0.000, 0.000]} & -- & -- \\
{\laya} & 0.564 & 0.618 & $-$0.053\,{\color{black!60}[$-$0.193, 0.083]} & -- & -- \\
{\deepseek} & \cellcolor{ebCoral!35}\textbf{1.000} & 1.000 & \cellcolor{ebCoral!35}\textbf{0.000}\,{\color{black!60}[0.000, 0.000]} & -- & -- \\
{\glm} & \cellcolor{ebCoral!35}\textbf{1.000} & 1.000 & \cellcolor{ebCoral!35}\textbf{0.000}\,{\color{black!60}[0.000, 0.000]} & -- & -- \\
{\qwenflash} & \cellcolor{ebCoral!35}\textbf{1.000} & 1.000 & \cellcolor{ebCoral!35}\textbf{0.000}\,{\color{black!60}[0.000, 0.000]} & -- & -- \\
{\qwenjson{} (ref.)} & 0.980 & 1.000 & $-$0.020\,{\color{black!60}[$-$0.041, $-$0.005]} & -- & -- \\
{MiniLM-Reranker} & 0.891 & 0.941 & $-$0.050\,{\color{black!60}[$-$0.117, 0.022]} & -- & -- \\
{DistilBERT-Clf-Frozen} & 0.653 & 0.000 & 0.653\,{\color{black!60}[0.586, 0.717]} & 0 & -- \\
{DistilBERT-Clf-Retrained} & 0.708 & 0.147 & 0.561\,{\color{black!60}[0.451, 0.663]} & 76 & 12.8 \\
\rowcolor{ebAmber!45}\multicolumn{6}{l}{\textbf{\textit{50\% churn}}} \\
\rowcolor{ebCoral!10}{\jev} & \cellcolor{ebCoral!35}\textbf{1.000} & \cellcolor{ebCoral!35}\textbf{1.000} & \cellcolor{ebCoral!35}\textbf{0.000}\,{\color{black!60}[0.000, 0.000]} & -- & -- \\
\cellcolor{ebPink}{\semif}$^\dagger$ & \cellcolor{ebPink}0.000 & \cellcolor{ebPink}0.000 & \cellcolor{ebPink}0.000\,{\color{black!60}[0.000, 0.000]} & -- & -- \\
{\laya} & 0.588 & 0.672 & $-$0.083\,{\color{black!60}[$-$0.198, 0.031]} & -- & -- \\
{\deepseek} & \cellcolor{ebCoral!35}\textbf{1.000} & \cellcolor{ebCoral!35}\textbf{1.000} & \cellcolor{ebCoral!35}\textbf{0.000}\,{\color{black!60}[0.000, 0.000]} & -- & -- \\
{\glm} & \cellcolor{ebCoral!35}\textbf{1.000} & \cellcolor{ebCoral!35}\textbf{1.000} & \cellcolor{ebCoral!35}\textbf{0.000}\,{\color{black!60}[0.000, 0.000]} & -- & -- \\
{\qwenflash} & \cellcolor{ebCoral!35}\textbf{1.000} & \cellcolor{ebCoral!35}\textbf{1.000} & \cellcolor{ebCoral!35}\textbf{0.000}\,{\color{black!60}[0.000, 0.000]} & -- & -- \\
{\qwenjson{} (ref.)} & 0.971 & 0.985 & $-$0.014\,{\color{black!60}[$-$0.053, 0.020]} & -- & -- \\
{MiniLM-Reranker} & 0.831 & 0.933 & $-$0.102\,{\color{black!60}[$-$0.179, $-$0.028]} & -- & -- \\
{DistilBERT-Clf-Frozen} & 0.522 & 0.000 & 0.522\,{\color{black!60}[0.439, 0.607]} & 0 & -- \\
{DistilBERT-Clf-Retrained} & 0.441 & 0.142 & 0.299\,{\color{black!60}[0.198, 0.402]} & 92 & 10.0 \\
\bottomrule
\end{tabular}}
\def\ebh#1#2{#1 #2}\sbox0{\ebtab}
\ifdim\wd0>\dimexpr\linewidth-1pt\relax\def\ebh#1#2{\shortstack{#1\\{}#2}}\sbox0{\ebtab}\fi
\setlength{\tabcolsep}{\dimexpr\tabcolsep+(\linewidth-\wd0-1pt)/12\relax}
\ifdim\tabcolsep<1.5pt\setlength{\tabcolsep}{1.5pt}\fi
\typeout{EBTAB tab_rq4b natural=\the\wd0\space line=\the\linewidth\space sep=\the\tabcolsep}
\ebtab
\par\smallskip{\scriptsize\ebflagnote\par}
\end{table}

Interpreters that receive the catalog with each request cope with both its size
and its churn. \jev and the three LLMs name known services with a top-1 accuracy
of at least 0.985 at every catalog size. These four interpreters name new
services with 0.995--1.000 at
128 and 254 services and under both churn levels (Fig.~\ref{fig:rq4} and
Table~\ref{tab:rq4a}). All 16 seen-minus-unseen contrasts stay below the 10-point bound (H5),
including those of \laya, \qwenjson, and the \reranker. The two \semif rows are
flagged because it returns no valid output there. \jev detects unsupported requests less reliably, with an
F1 score of 0.723--0.929 against 0.889--1.000 for the LLMs. Its median latency
stays at 0.266--0.335\,s from 4 to 254 services. Its fee grows with the number
of options, and from 64 services on it costs more per correct decision than
\deepseek.

Option limits bound the self-hosted decision models. \semif rejects every
request with more than 16 options, which removes it from the catalogs of 64 or
more services and from both churn conditions. \laya rejects the 254-service
catalog, and its accuracy falls from 0.981 with 4 services to 0.299 with 128.
Both limits are properties of the interfaces, recorded here as observed
failures.

The trained references in Table~\ref{tab:rq4b} show what request-time catalogs avoid. The frozen
\clf names no new service, and its accuracy on known services is 0.653 and
0.522 at 25\% and 50\% churn. Retraining with 76 and 92 labelled examples, 12.8
and 10.0\,s of training, lifts its unseen accuracy only to 0.147 and 0.142.
The \reranker names 0.941 and 0.933 of new services in 3\,ms, but its F1 score
for unsupported requests is 0.280 and 0.507.

\takeaway{Dynamic catalog}{Passing the catalog with the request lets \jev and
the LLMs absorb 50\% churn with no loss on unseen services, whereas a classifier
retrained on 92 labelled examples reaches 0.142. \semif stops at 16 options, \laya fails at 254, and \jev's fee
grows with the catalog. In each case the interface sets the limit.}

\subsection{Tail Latency, Jitter, and Energy}
\label{sec:comms}
\begin{figure}[!t]
\centering
\includegraphics[width=3.36in]{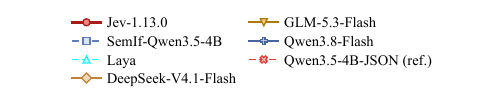}\\[-2pt]
\subfloat[Clean requests (RQ2)]{\includegraphics[width=1.68in]{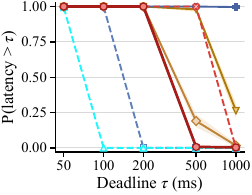}\label{fig:comms_latency_a}}\hfill
\subfloat[16{,}384-token inputs (RQ1a)]{\includegraphics[width=1.68in]{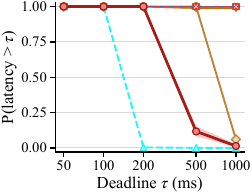}\label{fig:comms_latency_b}}
\caption{Decision-latency distributions as the probability that a decision exceeds a latency budget, (a) on clean requests and (b) on requests padded to 16{,}384 tokens.}
\label{fig:comms_latency}
\end{figure}

\begin{figure}[!t]
\centering
\includegraphics[width=3.36in]{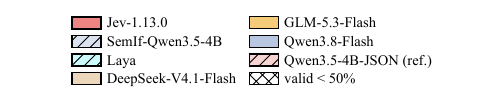}\\[-2pt]
\subfloat[Hosted: USD per 1k correct]{\includegraphics[width=1.68in]{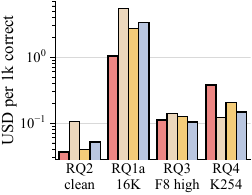}\label{fig:comms_efficiency_a}}\hfill
\subfloat[Self-hosted: J per decision]{\includegraphics[width=1.68in]{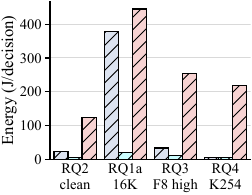}\label{fig:comms_efficiency_b}}
\caption{Efficiency per decision: (a) API fees per 1{,}000 correct decisions for the hosted interpreters and (b) accelerator energy per decision for the self-hosted ones, in four representative conditions.}
\label{fig:comms_efficiency}
\end{figure}

\begin{table*}[!t]
\centering
\caption{Communication-level metrics in four representative conditions: tail latency, jitter, probability of exceeding a latency budget, availability, goodput, SLA violations, payload bytes, API fees, and energy per decision.}
\label{tab:comms}
\scriptsize
\setlength{\tabcolsep}{3.2pt}
\setlength{\aboverulesep}{0pt}\setlength{\belowrulesep}{0pt}\setlength{\extrarowheight}{0pt}
\renewcommand{\arraystretch}{0.94}
\providecommand{\ebflagnote}{$^\dagger$Valid output rate below 0.5; the row is printed but excluded from the best-value marking.}
\def\ebtab{%
\begin{tabular}{lrrrrrrrrrrrr}
\toprule
\rowcolor{ebBlue}Model & \ebh{p95}{(s) $\downarrow$} & \ebh{p99}{(s) $\downarrow$} & \ebh{Jitter IQR}{(s) $\downarrow$} & \ebh{P($>$0.2\,s)}{$\downarrow$} & \ebh{P($>$1\,s)}{$\downarrow$} & \ebh{Avail.}{$\uparrow$} & \ebh{Goodput}{(corr./s) $\uparrow$} & \ebh{SLA}{violation $\downarrow$} & \ebh{Request}{(B) $\downarrow$} & \ebh{Response}{(B) $\downarrow$} & \ebh{USD per 1k}{correct $\downarrow$} & \ebh{Energy}{(J/dec.) $\downarrow$} \\
\midrule
\rowcolor{ebAmber!45}\multicolumn{13}{l}{\textbf{\textit{RQ2, clean}}} \\
\rowcolor{ebCoral!10}{\jev} & 0.370 & 0.438 & 0.064 & 1.000 & 0.003 & 1.000 & 3.33 & 0.003 & 186.5 & 728.5 & \cellcolor{ebCoral!35}\textbf{0.03612} & -- \\
{\semif} & 0.123 & 0.123 & 0.001 & \cellcolor{ebCoral!35}\textbf{0.000} & \cellcolor{ebCoral!35}\textbf{0.000} & 1.000 & \cellcolor{ebCoral!35}\textbf{5.27} & 0.080 & 186.5 & 1031 & -- & 22.5 \\
{\laya} & \cellcolor{ebCoral!35}\textbf{0.056} & \cellcolor{ebCoral!35}\textbf{0.056} & \cellcolor{ebCoral!35}\textbf{0.000} & \cellcolor{ebCoral!35}\textbf{0.000} & \cellcolor{ebCoral!35}\textbf{0.000} & 1.000 & 2.01 & 0.143 & 186.5 & 849 & -- & \cellcolor{ebCoral!35}\textbf{6.1} \\
{\deepseek} & 0.692 & 1.280 & 0.104 & 1.000 & 0.017 & 1.000 & 2.19 & 0.003 & 186.5 & 927 & 0.1065 & -- \\
{\glm} & 2.940 & 3.982 & 0.443 & 1.000 & 0.263 & 1.000 & 0.93 & 0.003 & 186.5 & 1115 & 0.03971 & -- \\
{\qwenflash} & 3.450 & 5.099 & 0.689 & 1.000 & 0.997 & 0.997 & 0.49 & 0.003 & 186.5 & 925 & 0.05172 & -- \\
{\qwenjson{} (ref.)} & 0.811 & 0.833 & 0.043 & 1.000 & \cellcolor{ebCoral!35}\textbf{0.000} & 1.000 & 0.63 & 0.053 & 186.5 & \cellcolor{ebCoral!35}\textbf{511} & -- & 123.3 \\
\rowcolor{ebAmber!45}\multicolumn{13}{l}{\textbf{\textit{RQ1a, 16{,}384-token input}}} \\
\rowcolor{ebCoral!10}{\jev} & 0.523 & 1.198 & 0.055 & 1.000 & 0.013 & 1.000 & \cellcolor{ebCoral!35}\textbf{1.89} & 0.013 & 41749 & 740 & \cellcolor{ebCoral!35}\textbf{1.044} & -- \\
{\semif} & 1.238 & 1.288 & 0.016 & 1.000 & 1.000 & 1.000 & 0.41 & 0.340 & 41749 & 1033 & -- & 378.8 \\
{\laya} & \cellcolor{ebCoral!35}\textbf{0.140} & \cellcolor{ebCoral!35}\textbf{0.142} & \cellcolor{ebCoral!35}\textbf{0.002} & \cellcolor{ebCoral!35}\textbf{0.003} & \cellcolor{ebCoral!35}\textbf{0.000} & 1.000 & 0.12 & 0.500 & 41749 & 852 & -- & \cellcolor{ebCoral!35}\textbf{20.4} \\
{\deepseek} & 1.023 & 1.183 & 0.170 & 1.000 & 0.060 & 1.000 & 1.40 & 0.003 & 41749 & 937 & 5.455 & -- \\
{\glm} & 3.939 & 12.754 & 0.829 & 1.000 & 0.990 & 0.980 & 0.44 & 0.050 & 41749 & 937 & 2.727 & -- \\
{\qwenflash} & 4.274 & 5.400 & 0.670 & 1.000 & 1.000 & 1.000 & 0.31 & \cellcolor{ebCoral!35}\textbf{0.000} & 41749 & 945 & 3.373 & -- \\
{\qwenjson{} (ref.)} & 1.619 & 2.384 & 0.024 & 1.000 & 1.000 & 1.000 & 0.11 & 0.127 & 41749 & \cellcolor{ebCoral!35}\textbf{514} & -- & 444.7 \\
\rowcolor{ebAmber!45}\multicolumn{13}{l}{\textbf{\textit{RQ3, $F=8$, high density}}} \\
\rowcolor{ebCoral!10}{\jev} & 0.350 & 0.401 & 0.047 & 1.000 & \cellcolor{ebCoral!35}\textbf{0.000} & 1.000 & \cellcolor{ebCoral!35}\textbf{1.91} & 0.057 & 402.5 & 1313 & 0.1119 & -- \\
{\semif} & 0.221 & 0.267 & 0.040 & 0.080 & \cellcolor{ebCoral!35}\textbf{0.000} & 1.000 & 0.47 & 0.223 & 402.5 & 1935 & -- & 33.6 \\
{\laya} & \cellcolor{ebCoral!35}\textbf{0.193} & \cellcolor{ebCoral!35}\textbf{0.224} & \cellcolor{ebCoral!35}\textbf{0.021} & \cellcolor{ebCoral!35}\textbf{0.040} & \cellcolor{ebCoral!35}\textbf{0.000} & 1.000 & 0.04 & 0.533 & 402.5 & 1544 & -- & \cellcolor{ebCoral!35}\textbf{10.7} \\
{\deepseek} & 0.953 & 7.556 & 0.091 & 1.000 & 0.050 & 1.000 & 1.45 & \cellcolor{ebCoral!35}\textbf{0.010} & 402.5 & 1059 & 0.1410 & -- \\
{\glm} & 2.462 & 4.848 & 0.449 & 1.000 & 0.843 & 1.000 & 0.60 & 0.100 & 402.5 & 1405.5 & 0.1245 & -- \\
{\qwenflash} & 5.592 & 7.078 & 0.517 & 1.000 & 1.000 & 0.960 & 0.27 & 0.050 & 402.5 & 1079.5 & \cellcolor{ebCoral!35}\textbf{0.1037} & -- \\
{\qwenjson{} (ref.)} & 1.657 & 1.951 & 0.074 & 1.000 & 1.000 & 1.000 & 0.13 & 0.113 & 402.5 & \cellcolor{ebCoral!35}\textbf{636.5} & -- & 254.4 \\
\rowcolor{ebAmber!45}\multicolumn{13}{l}{\textbf{\textit{RQ4, $K=254$ services}}} \\
\rowcolor{ebCoral!10}{\jev} & \cellcolor{ebCoral!35}\textbf{0.432} & \cellcolor{ebCoral!35}\textbf{0.502} & \cellcolor{ebCoral!35}\textbf{0.046} & 1.000 & \cellcolor{ebCoral!35}\textbf{0.003} & 1.000 & \cellcolor{ebCoral!35}\textbf{2.65} & 0.013 & 245.5 & 8161 & 0.3798 & -- \\
\cellcolor{ebPink}{\semif}$^\dagger$ & \cellcolor{ebPink}0.041 & \cellcolor{ebPink}0.041 & \cellcolor{ebPink}0.000 & \cellcolor{ebPink}0.000 & \cellcolor{ebPink}0.000 & \cellcolor{ebPink}0.000 & \cellcolor{ebPink}0.00 & \cellcolor{ebPink}1.000 & \cellcolor{ebPink}245.5 & \cellcolor{ebPink}128 & -- & \cellcolor{ebPink}4.8 \\
\cellcolor{ebPink}{\laya}$^\dagger$ & \cellcolor{ebPink}0.052 & \cellcolor{ebPink}0.053 & \cellcolor{ebPink}0.000 & \cellcolor{ebPink}0.000 & \cellcolor{ebPink}0.000 & \cellcolor{ebPink}0.000 & \cellcolor{ebPink}0.00 & \cellcolor{ebPink}1.000 & \cellcolor{ebPink}245.5 & \cellcolor{ebPink}119 & -- & \cellcolor{ebPink}4.9 \\
{\deepseek} & 1.008 & 1.574 & 0.090 & 1.000 & 0.053 & 1.000 & 1.78 & \cellcolor{ebCoral!35}\textbf{0.003} & 245.5 & 965 & \cellcolor{ebCoral!35}\textbf{0.1206} & -- \\
{\glm} & 1.876 & 2.699 & 0.202 & 1.000 & 0.163 & 1.000 & 1.13 & 0.017 & 245.5 & 1153 & 0.2046 & -- \\
{\qwenflash} & 3.791 & 5.274 & 0.656 & 1.000 & 1.000 & 1.000 & 0.47 & 0.007 & 245.5 & 957 & 0.1473 & -- \\
{\qwenjson{} (ref.)} & 1.179 & 1.671 & 0.075 & 1.000 & 0.867 & 1.000 & 0.45 & 0.053 & 245.5 & \cellcolor{ebCoral!35}\textbf{529} & -- & 218.7 \\
\bottomrule
\end{tabular}}
\def\ebh#1#2{#1 #2}\sbox0{\ebtab}
\ifdim\wd0>\dimexpr\linewidth-1pt\relax\def\ebh#1#2{\shortstack{#1\\{}#2}}\sbox0{\ebtab}\fi
\setlength{\tabcolsep}{\dimexpr\tabcolsep+(\linewidth-\wd0-1pt)/26\relax}
\ifdim\tabcolsep<1.5pt\setlength{\tabcolsep}{1.5pt}\fi
\typeout{EBTAB tab_comms natural=\the\wd0\space line=\the\linewidth\space sep=\the\tabcolsep}
\ebtab
\par\smallskip{\scriptsize\ebflagnote\par}
\end{table*}

Tail behavior follows the same order as the medians. On clean requests, \jev
exceeds 0.5\,s in 0.7\% of decisions, against 19.0\% for \deepseek, 98.0\% for
\glm, and 100\% for \qwenflash (Fig.~\ref{fig:comms_latency} and Table~\ref{tab:comms}). Its interquartile range
is 64\,ms, against 104, 443, and 689\,ms. The self-hosted decision models
answer within 0.5\,s every time and use 22.5\,J (\semif) and 6.1\,J (\laya)
per decision, against 123.3\,J for \qwenjson on the same accelerator
(Fig.~\ref{fig:comms_efficiency}).

\begin{table*}[!t]
\centering
\caption{Pre-registered confirmatory contrasts H1--H5 (the paired H3 contrasts are in Table~\ref{tab:rq2_contrasts}). A contrast is resolved when its 95\% confidence interval excludes the null value and its Holm-adjusted $p$ is below 0.05.}
\label{tab:hypotheses}
\scriptsize
\setlength{\tabcolsep}{6pt}
\setlength{\aboverulesep}{0pt}\setlength{\belowrulesep}{0pt}\setlength{\extrarowheight}{0pt}
\renewcommand{\arraystretch}{0.94}
\providecommand{\ebflagnote}{$^\dagger$Valid output rate below 0.5; the row is printed but excluded from the best-value marking.}
\def\ebtab{%
\begin{tabular}{p{3.05in}lrrl}
\toprule
\rowcolor{ebBlue}Contrast & Level & Estimate [95\% CI] & Holm $p$ & Verdict \\
\midrule
\rowcolor{ebAmber!45}\multicolumn{5}{l}{\textbf{\textit{H1(a): paired p50 latency difference (s)}}} \\
DeepSeek-V4.1-Flash $-$ Jev-1.13.0 & Base length & 0.109\,{\color{black!60}[0.102, 0.117]} & 3.8e-46 & \cellcolor{ebCoral!35}\textbf{resolved} \\
DeepSeek-V4.1-Flash $-$ Jev-1.13.0 & Padded to 512 tokens & 0.217\,{\color{black!60}[0.204, 0.231]} & 5.8e-48 & \cellcolor{ebCoral!35}\textbf{resolved} \\
DeepSeek-V4.1-Flash $-$ Jev-1.13.0 & Padded to 2{,}048 tokens & 0.269\,{\color{black!60}[0.257, 0.283]} & 6.7e-49 & \cellcolor{ebCoral!35}\textbf{resolved} \\
DeepSeek-V4.1-Flash $-$ Jev-1.13.0 & Padded to 8{,}192 tokens & 0.165\,{\color{black!60}[0.151, 0.183]} & 6.8e-42 & \cellcolor{ebCoral!35}\textbf{resolved} \\
DeepSeek-V4.1-Flash $-$ Jev-1.13.0 & Padded to 16{,}384 tokens & 0.218\,{\color{black!60}[0.204, 0.236]} & 1.2e-45 & \cellcolor{ebCoral!35}\textbf{resolved} \\
GLM-5.3-Flash $-$ Jev-1.13.0 & Base length & 0.370\,{\color{black!60}[0.347, 0.396]} & 4.9e-49 & \cellcolor{ebCoral!35}\textbf{resolved} \\
GLM-5.3-Flash $-$ Jev-1.13.0 & Padded to 512 tokens & 0.447\,{\color{black!60}[0.425, 0.470]} & 1.2e-49 & \cellcolor{ebCoral!35}\textbf{resolved} \\
GLM-5.3-Flash $-$ Jev-1.13.0 & Padded to 2{,}048 tokens & 0.400\,{\color{black!60}[0.370, 0.427]} & 1.2e-49 & \cellcolor{ebCoral!35}\textbf{resolved} \\
GLM-5.3-Flash $-$ Jev-1.13.0 & Padded to 8{,}192 tokens & 0.751\,{\color{black!60}[0.722, 0.811]} & 6.2e-49 & \cellcolor{ebCoral!35}\textbf{resolved} \\
GLM-5.3-Flash $-$ Jev-1.13.0 & Padded to 16{,}384 tokens & 1.156\,{\color{black!60}[1.072, 1.218]} & 6.7e-47 & \cellcolor{ebCoral!35}\textbf{resolved} \\
Qwen3.8-Flash $-$ Jev-1.13.0 & Base length & 1.472\,{\color{black!60}[1.379, 1.616]} & 1.2e-49 & \cellcolor{ebCoral!35}\textbf{resolved} \\
Qwen3.8-Flash $-$ Jev-1.13.0 & Padded to 512 tokens & 1.358\,{\color{black!60}[1.294, 1.459]} & 1.2e-49 & \cellcolor{ebCoral!35}\textbf{resolved} \\
Qwen3.8-Flash $-$ Jev-1.13.0 & Padded to 2{,}048 tokens & 1.314\,{\color{black!60}[1.274, 1.349]} & 1.2e-49 & \cellcolor{ebCoral!35}\textbf{resolved} \\
Qwen3.8-Flash $-$ Jev-1.13.0 & Padded to 8{,}192 tokens & 1.994\,{\color{black!60}[1.946, 2.066]} & 1.2e-49 & \cellcolor{ebCoral!35}\textbf{resolved} \\
Qwen3.8-Flash $-$ Jev-1.13.0 & Padded to 16{,}384 tokens & 2.497\,{\color{black!60}[2.416, 2.553]} & 5.8e-48 & \cellcolor{ebCoral!35}\textbf{resolved} \\
Qwen3.5-4B-JSON $-$ SemIf-Qwen3.5-4B & Base length & 0.648\,{\color{black!60}[0.635, 0.649]} & 1.2e-49 & \cellcolor{ebCoral!35}\textbf{resolved} \\
Qwen3.5-4B-JSON $-$ SemIf-Qwen3.5-4B & Padded to 512 tokens & 0.637\,{\color{black!60}[0.633, 0.646]} & 1.2e-49 & \cellcolor{ebCoral!35}\textbf{resolved} \\
Qwen3.5-4B-JSON $-$ SemIf-Qwen3.5-4B & Padded to 2{,}048 tokens & 0.645\,{\color{black!60}[0.643, 0.646]} & 1.2e-49 & \cellcolor{ebCoral!35}\textbf{resolved} \\
Qwen3.5-4B-JSON $-$ SemIf-Qwen3.5-4B & Padded to 8{,}192 tokens & 0.534\,{\color{black!60}[0.532, 0.537]} & 1.2e-49 & \cellcolor{ebCoral!35}\textbf{resolved} \\
Qwen3.5-4B-JSON $-$ SemIf-Qwen3.5-4B & Padded to 16{,}384 tokens & 0.393\,{\color{black!60}[0.391, 0.395]} & 1.2e-49 & \cellcolor{ebCoral!35}\textbf{resolved} \\
\rowcolor{ebAmber!45}\multicolumn{5}{l}{\textbf{\textit{H1(b): ratio of latency growth factors}}} \\
\rowcolor{ebCoral!10}$G$(Jev-1.13.0; 16K) / $G$(DeepSeek-V4.1-Flash; 16K) & 16{,}384 tokens vs base & 0.904\,{\color{black!60}[0.865, 0.940]} & $\le$4.0e-4 & \cellcolor{ebCoral!35}\textbf{resolved} \\
\rowcolor{ebCoral!10}$G$(Jev-1.13.0; 16K) / $G$(GLM-5.3-Flash; 16K) & 16{,}384 tokens vs base & 0.622\,{\color{black!60}[0.585, 0.663]} & $\le$4.0e-4 & \cellcolor{ebCoral!35}\textbf{resolved} \\
\rowcolor{ebCoral!10}$G$(Jev-1.13.0; 16K) / $G$(Qwen3.8-Flash; 16K) & 16{,}384 tokens vs base & 0.929\,{\color{black!60}[0.866, 0.986]} & 0.018 & \cellcolor{ebCoral!35}\textbf{resolved} \\
$G$(SemIf-Qwen3.5-4B; 16K) / $G$(Qwen3.5-4B-JSON; 16K) & 16{,}384 tokens vs base & 4.675\,{\color{black!60}[4.665, 4.701]} & $\le$4.0e-4 & not resolved \\
\rowcolor{ebAmber!45}\multicolumn{5}{l}{\textbf{\textit{H2: ratio of latency growth factors}}} \\
\rowcolor{ebCoral!10}$R$(Jev-1.13.0; 8) / $R$(DeepSeek-V4.1-Flash; 8) & $k=8$ vs $k=1$ & 0.494\,{\color{black!60}[0.468, 0.519]} & $\le$4.0e-4 & \cellcolor{ebCoral!35}\textbf{resolved} \\
\rowcolor{ebCoral!10}$R$(Jev-1.13.0; 8) / $R$(GLM-5.3-Flash; 8) & $k=8$ vs $k=1$ & 0.399\,{\color{black!60}[0.379, 0.427]} & $\le$4.0e-4 & \cellcolor{ebCoral!35}\textbf{resolved} \\
\rowcolor{ebCoral!10}$R$(Jev-1.13.0; 8) / $R$(Qwen3.8-Flash; 8) & $k=8$ vs $k=1$ & 0.367\,{\color{black!60}[0.350, 0.387]} & $\le$4.0e-4 & \cellcolor{ebCoral!35}\textbf{resolved} \\
$R$(SemIf-Qwen3.5-4B; 8) / $R$(Qwen3.5-4B-JSON; 8) & $k=8$ vs $k=1$ & 0.528\,{\color{black!60}[0.515, 0.575]} & $\le$4.0e-4 & \cellcolor{ebCoral!35}\textbf{resolved} \\
\rowcolor{ebAmber!45}\multicolumn{5}{l}{\textbf{\textit{H3: Cochran's $Q$ on unsafe decisions (statistic)}}} \\
Cochran's Q on unsafe across six & Clean & 188.3 & 7.2e-38 & \cellcolor{ebCoral!35}\textbf{resolved} \\
Cochran's Q on unsafe across six & Code-switched & 171.9 & 1.7e-34 & \cellcolor{ebCoral!35}\textbf{resolved} \\
Cochran's Q on unsafe across six & Colloquial & 98.0 & 4.2e-19 & \cellcolor{ebCoral!35}\textbf{resolved} \\
Cochran's Q on unsafe across six & Default-baiting & 82.7 & 4.6e-16 & \cellcolor{ebCoral!35}\textbf{resolved} \\
Cochran's Q on unsafe across six & Key--value & 50.0 & 1.4e-9 & \cellcolor{ebCoral!35}\textbf{resolved} \\
Cochran's Q on unsafe across six & Negation-heavy & 120.3 & 1.1e-23 & \cellcolor{ebCoral!35}\textbf{resolved} \\
Cochran's Q on unsafe across six & Noisy & 171.0 & 2.2e-34 & \cellcolor{ebCoral!35}\textbf{resolved} \\
Cochran's Q on unsafe across six & Self-revising & 172.2 & 1.7e-34 & \cellcolor{ebCoral!35}\textbf{resolved} \\
\rowcolor{ebAmber!45}\multicolumn{5}{l}{\textbf{\textit{H4: paired output-token difference (tokens)}}} \\
DeepSeek-V4.1-Flash output tokens F8 $-$ F4 & $F=8$ vs $F=4$, all $D$ & 28.0\,{\color{black!60}[27.0, 28.0]} & $\le$3.0e-4 & \cellcolor{ebCoral!35}\textbf{resolved} \\
GLM-5.3-Flash output tokens F8 $-$ F4 & $F=8$ vs $F=4$, all $D$ & 36.0\,{\color{black!60}[35.0, 37.0]} & $\le$3.0e-4 & \cellcolor{ebCoral!35}\textbf{resolved} \\
Qwen3.8-Flash output tokens F8 $-$ F4 & $F=8$ vs $F=4$, all $D$ & 43.0\,{\color{black!60}[42.0, 44.5]} & $\le$3.0e-4 & \cellcolor{ebCoral!35}\textbf{resolved} \\
\rowcolor{ebAmber!45}\multicolumn{5}{l}{\textbf{\textit{H4: ratio of latency growth factors}}} \\
\rowcolor{ebCoral!10}$G$(Jev-1.13.0; F8/F4) / $G$(DeepSeek-V4.1-Flash; F8/F4) & $F=8$ vs $F=4$, all $D$ & 0.828\,{\color{black!60}[0.807, 0.844]} & $\le$4.0e-4 & \cellcolor{ebCoral!35}\textbf{resolved} \\
\rowcolor{ebCoral!10}$G$(Jev-1.13.0; F8/F4) / $G$(GLM-5.3-Flash; F8/F4) & $F=8$ vs $F=4$, all $D$ & 0.638\,{\color{black!60}[0.620, 0.652]} & $\le$4.0e-4 & \cellcolor{ebCoral!35}\textbf{resolved} \\
\rowcolor{ebCoral!10}$G$(Jev-1.13.0; F8/F4) / $G$(Qwen3.8-Flash; F8/F4) & $F=8$ vs $F=4$, all $D$ & 0.650\,{\color{black!60}[0.633, 0.670]} & $\le$4.0e-4 & \cellcolor{ebCoral!35}\textbf{resolved} \\
$G$(SemIf-Qwen3.5-4B; F8/F4) / $G$(Qwen3.5-4B-JSON; F8/F4) & $F=8$ vs $F=4$, all $D$ & 0.561\,{\color{black!60}[0.559, 0.562]} & $\le$4.0e-4 & \cellcolor{ebCoral!35}\textbf{resolved} \\
\rowcolor{ebAmber!45}\multicolumn{5}{l}{\textbf{\textit{H5: seen $-$ unseen top-1 gap (proportion)}}} \\
\rowcolor{ebCoral!10}Jev-1.13.0 seen $-$ unseen gap & 25\% churn & 0.000\,{\color{black!60}[0.000, 0.000]} & $\le$0.0016 & \cellcolor{ebCoral!35}\textbf{resolved} \\
DeepSeek-V4.1-Flash seen $-$ unseen gap & 25\% churn & 0.000\,{\color{black!60}[0.000, 0.000]} & $\le$0.0016 & \cellcolor{ebCoral!35}\textbf{resolved} \\
GLM-5.3-Flash seen $-$ unseen gap & 25\% churn & 0.000\,{\color{black!60}[0.000, 0.000]} & $\le$0.0016 & \cellcolor{ebCoral!35}\textbf{resolved} \\
Qwen3.8-Flash seen $-$ unseen gap & 25\% churn & 0.000\,{\color{black!60}[0.000, 0.000]} & $\le$0.0016 & \cellcolor{ebCoral!35}\textbf{resolved} \\
\cellcolor{ebPink}SemIf-Qwen3.5-4B seen $-$ unseen gap$^\dagger$ & \cellcolor{ebPink}25\% churn & \cellcolor{ebPink}0.000\,{\color{black!60}[0.000, 0.000]} & \cellcolor{ebPink}$\le$0.0016 & \cellcolor{ebPink}resolved \\
Laya seen $-$ unseen gap & 25\% churn & $-$0.053\,{\color{black!60}[$-$0.193, 0.083]} & 0.027 & \cellcolor{ebCoral!35}\textbf{resolved} \\
Qwen3.5-4B-JSON seen $-$ unseen gap & 25\% churn & $-$0.020\,{\color{black!60}[$-$0.041, $-$0.005]} & $\le$0.0016 & \cellcolor{ebCoral!35}\textbf{resolved} \\
MiniLM-Reranker seen $-$ unseen gap & 25\% churn & $-$0.050\,{\color{black!60}[$-$0.117, 0.022]} & $\le$0.0016 & \cellcolor{ebCoral!35}\textbf{resolved} \\
\rowcolor{ebCoral!10}Jev-1.13.0 seen $-$ unseen gap & 50\% churn & 0.000\,{\color{black!60}[0.000, 0.000]} & $\le$0.0016 & \cellcolor{ebCoral!35}\textbf{resolved} \\
DeepSeek-V4.1-Flash seen $-$ unseen gap & 50\% churn & 0.000\,{\color{black!60}[0.000, 0.000]} & $\le$0.0016 & \cellcolor{ebCoral!35}\textbf{resolved} \\
GLM-5.3-Flash seen $-$ unseen gap & 50\% churn & 0.000\,{\color{black!60}[0.000, 0.000]} & $\le$0.0016 & \cellcolor{ebCoral!35}\textbf{resolved} \\
Qwen3.8-Flash seen $-$ unseen gap & 50\% churn & 0.000\,{\color{black!60}[0.000, 0.000]} & $\le$0.0016 & \cellcolor{ebCoral!35}\textbf{resolved} \\
\cellcolor{ebPink}SemIf-Qwen3.5-4B seen $-$ unseen gap$^\dagger$ & \cellcolor{ebPink}50\% churn & \cellcolor{ebPink}0.000\,{\color{black!60}[0.000, 0.000]} & \cellcolor{ebPink}$\le$0.0016 & \cellcolor{ebPink}resolved \\
Laya seen $-$ unseen gap & 50\% churn & $-$0.083\,{\color{black!60}[$-$0.198, 0.031]} & 0.0016 & \cellcolor{ebCoral!35}\textbf{resolved} \\
Qwen3.5-4B-JSON seen $-$ unseen gap & 50\% churn & $-$0.014\,{\color{black!60}[$-$0.053, 0.020]} & $\le$0.0016 & \cellcolor{ebCoral!35}\textbf{resolved} \\
MiniLM-Reranker seen $-$ unseen gap & 50\% churn & $-$0.102\,{\color{black!60}[$-$0.179, $-$0.028]} & $\le$0.0016 & \cellcolor{ebCoral!35}\textbf{resolved} \\
\bottomrule
\end{tabular}}
\def\ebh#1#2{#1 #2}\sbox0{\ebtab}
\ifdim\wd0>\dimexpr\linewidth-1pt\relax\def\ebh#1#2{\shortstack{#1\\{}#2}}\sbox0{\ebtab}\fi
\setlength{\tabcolsep}{\dimexpr\tabcolsep+(\linewidth-\wd0-1pt)/10\relax}
\ifdim\tabcolsep<1.5pt\setlength{\tabcolsep}{1.5pt}\fi
\typeout{EBTAB tab_hypotheses natural=\the\wd0\space line=\the\linewidth\space sep=\the\tabcolsep}
\ebtab
\par\smallskip{\scriptsize\ebflagnote\par}
\end{table*}

\subsection{RQ5: End-to-End Service Outcome}
\label{sec:rq5}
\begin{figure*}[!t]
\centering
\includegraphics[width=7.00in]{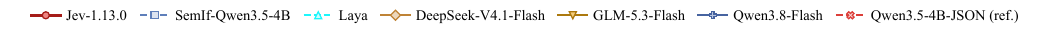}\\[-2pt]
\subfloat[Completion vs.\ $\lambda$]{\includegraphics[width=1.70in]{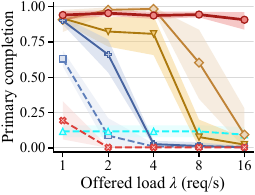}\label{fig:rq5a_a}}\hfill
\subfloat[Latency p95 vs.\ $\lambda$]{\includegraphics[width=1.70in]{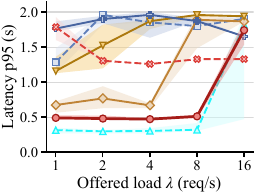}\label{fig:rq5a_b}}\hfill
\subfloat[Completion vs.\ deadline $D$]{\includegraphics[width=1.70in]{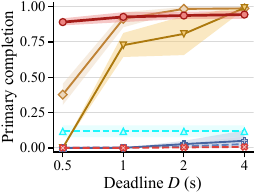}\label{fig:rq5a_c}}\hfill
\subfloat[Cache reuse block]{\includegraphics[width=1.70in]{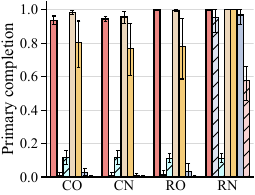}\label{fig:rq5a_d}}
\caption{RQ5, modeled execution (Part~A). (a) Primary completion and (b) 95th percentile of full request latency $T$ on completions as offered load $\lambda$ increases from 1 to 16 requests/s. (c) Primary completion across request deadline $D$ from 0.5 to 4~s ($D=2$~s is the $\lambda=4$ cell). Bands are 95\% bootstrap confidence intervals. (d) Primary completion under cache reuse conditions: CO changing text with cache off, CN changing text with cache on, RO repeated text with cache off, RN repeated text with cache on.}
\label{fig:rq5a}
\end{figure*}

\begin{table*}[!t]
\centering
\caption{RQ5, modeled execution (Part~A). Primary completion rate and 95th percentile request latency $T$ on completions across offered load, deadline, topology, cache reuse, and burstiness sweeps. The $\lambda=4$ cell serves as the reference for $D=2$\,s, $N=5$, and changing text with cache off. Model codes: JEV \jev, SIF \semif, LAY \laya, DSK \deepseek, GLM \glm, Q38 \qwenflash, QJS \qwenjson{} (ref.). Best value per row within each metric block in bold (QJS reference excluded). 95\% confidence intervals are drawn in Fig.~\ref{fig:rq5a}.}
\label{tab:rq5a}
\scriptsize
\setlength{\tabcolsep}{3pt}
\setlength{\aboverulesep}{0pt}\setlength{\belowrulesep}{0pt}\setlength{\extrarowheight}{0pt}
\renewcommand{\arraystretch}{0.94}
\def\ebtab{%
\begin{tabular}{lrrrrrrrrrrrrrr}
\toprule
\rowcolor{ebBlue} & \multicolumn{7}{c}{Primary completion $\uparrow$} & \multicolumn{7}{c}{p95 $T$ (s) $\downarrow$} \\
\cmidrule(lr){2-8}\cmidrule(lr){9-15}
\rowcolor{ebBlue}Cell & JEV & SIF & LAY & DSK & GLM & Q38 & QJS & JEV & SIF & LAY & DSK & GLM & Q38 & QJS \\
\midrule
\rowcolor{ebAmber!45}\multicolumn{15}{l}{\textbf{\textit{Offered load sweep ($\lambda \in \{1, 2, 4, 8, 16\}$ req/s)}}} \\
$\lambda{=}1$ & \cellcolor{ebCoral!35}\textbf{0.940} & 0.630 & 0.117 & 0.907 & 0.917 & 0.900 & 0.193 & \cellcolor{ebCoral!10}0.49 & 1.28 & \cellcolor{ebCoral!35}\textbf{0.31} & 0.67 & 1.16 & 1.76 & 1.78 \\
$\lambda{=}2$ & \cellcolor{ebCoral!10}0.953 & 0.087 & 0.117 & \cellcolor{ebCoral!35}\textbf{0.977} & 0.823 & 0.660 & 0.003 & \cellcolor{ebCoral!10}0.48 & 1.96 & \cellcolor{ebCoral!35}\textbf{0.30} & 0.77 & 1.52 & 1.90 & 1.30 \\
$\lambda{=}4$ & \cellcolor{ebCoral!10}0.937 & 0.010 & 0.117 & \cellcolor{ebCoral!35}\textbf{0.983} & 0.807 & 0.027 & 0.003 & \cellcolor{ebCoral!10}0.47 & 1.85 & \cellcolor{ebCoral!35}\textbf{0.30} & 0.67 & 1.87 & 1.96 & 1.26 \\
$\lambda{=}8$ & \cellcolor{ebCoral!35}\textbf{0.943} & 0.010 & 0.117 & 0.603 & 0.077 & 0.013 & 0.003 & \cellcolor{ebCoral!10}0.51 & 1.80 & \cellcolor{ebCoral!35}\textbf{0.32} & 1.89 & 1.96 & 1.87 & 1.33 \\
$\lambda{=}16$ & \cellcolor{ebCoral!35}\textbf{0.907} & 0.010 & 0.093 & 0.093 & 0.023 & 0.003 & 0.003 & \cellcolor{ebCoral!10}1.74 & 1.90 & 1.82 & 1.87 & 1.94 & \cellcolor{ebCoral!35}\textbf{1.65} & 1.33 \\
\rowcolor{ebAmber!45}\multicolumn{15}{l}{\textbf{\textit{Deadline sweep ($D \in \{0.5, 1, 4\}$ s; $D=2$\,s in load sweep)}}} \\
$D{=}0.5$\,s & \cellcolor{ebCoral!35}\textbf{0.890} & 0.000 & 0.117 & 0.377 & 0.000 & 0.000 & 0.000 & \cellcolor{ebCoral!10}0.46 & -- & \cellcolor{ebCoral!35}\textbf{0.29} & 0.49 & -- & -- & -- \\
$D{=}1$\,s & \cellcolor{ebCoral!35}\textbf{0.927} & 0.003 & 0.117 & 0.913 & 0.727 & 0.000 & 0.000 & \cellcolor{ebCoral!10}0.47 & 0.89 & \cellcolor{ebCoral!35}\textbf{0.29} & 0.83 & 0.97 & -- & -- \\
$D{=}4$\,s & \cellcolor{ebCoral!10}0.943 & 0.027 & 0.117 & 0.987 & \cellcolor{ebCoral!35}\textbf{0.990} & 0.050 & 0.007 & \cellcolor{ebCoral!10}0.48 & 3.83 & \cellcolor{ebCoral!35}\textbf{0.30} & 1.01 & 1.45 & 3.95 & 2.89 \\
\rowcolor{ebAmber!45}\multicolumn{15}{l}{\textbf{\textit{Topology sweep ($N \in \{10, 20, 40\}$ nodes; $N=5$ in load sweep)}}} \\
$N{=}10$ & \cellcolor{ebCoral!10}0.940 & 0.010 & 0.117 & \cellcolor{ebCoral!35}\textbf{0.993} & 0.023 & 0.013 & 0.003 & \cellcolor{ebCoral!10}0.48 & 1.82 & \cellcolor{ebCoral!35}\textbf{0.35} & 1.48 & 1.92 & 1.92 & 1.28 \\
$N{=}20$ & \cellcolor{ebCoral!35}\textbf{0.943} & 0.010 & 0.117 & 0.853 & 0.123 & 0.010 & 0.003 & \cellcolor{ebCoral!10}0.50 & 1.84 & \cellcolor{ebCoral!35}\textbf{0.34} & 1.85 & 1.99 & 1.71 & 1.26 \\
$N{=}40$ & \cellcolor{ebCoral!10}0.947 & 0.010 & 0.117 & \cellcolor{ebCoral!35}\textbf{0.990} & 0.023 & 0.007 & 0.003 & \cellcolor{ebCoral!10}0.51 & 1.84 & \cellcolor{ebCoral!35}\textbf{0.32} & 1.39 & 1.91 & 1.64 & 1.38 \\
\rowcolor{ebAmber!45}\multicolumn{15}{l}{\textbf{\textit{Cache reuse sweep ($\lambda=4$, $D=2$\,s; CO in load sweep)}}} \\
changing, on & \cellcolor{ebCoral!10}0.943 & 0.010 & 0.117 & \cellcolor{ebCoral!35}\textbf{0.957} & 0.770 & 0.007 & 0.003 & \cellcolor{ebCoral!10}0.47 & 1.65 & \cellcolor{ebCoral!35}\textbf{0.30} & 0.80 & 1.79 & 1.93 & 1.32 \\
repeated, off & \cellcolor{ebCoral!35}\textbf{1.000} & 0.013 & 0.113 & 0.997 & 0.780 & 0.033 & 0.003 & \cellcolor{ebCoral!10}0.48 & 1.89 & \cellcolor{ebCoral!35}\textbf{0.30} & 0.70 & 1.83 & 1.86 & 1.51 \\
repeated, on & \cellcolor{ebCoral!35}\textbf{1.000} & 0.953 & 0.113 & \cellcolor{ebCoral!35}\textbf{1.000} & \cellcolor{ebCoral!35}\textbf{1.000} & 0.970 & 0.577 & \cellcolor{ebCoral!10}0.20 & \cellcolor{ebCoral!35}\textbf{0.19} & 0.20 & 0.20 & 0.20 & \cellcolor{ebCoral!35}\textbf{0.19} & 0.19 \\
\rowcolor{ebAmber!45}\multicolumn{15}{l}{\textbf{\textit{Burstiness sweep ($\lambda=4$, $D=2$\,s, Poisson bursts)}}} \\
bursty & \cellcolor{ebCoral!10}0.950 & 0.083 & 0.117 & \cellcolor{ebCoral!35}\textbf{0.987} & 0.223 & 0.063 & 0.053 & \cellcolor{ebCoral!10}0.51 & 1.49 & \cellcolor{ebCoral!35}\textbf{0.33} & 0.97 & 1.93 & 1.93 & 1.77 \\
\bottomrule
\end{tabular}}
\def\ebh#1#2{#1 #2}\sbox0{\ebtab}
\ifdim\wd0>\dimexpr\linewidth-1pt\relax\def\ebh#1#2{\shortstack{#1\\{}#2}}\sbox0{\ebtab}\fi
\setlength{\tabcolsep}{\dimexpr\tabcolsep+(\linewidth-\wd0-1pt)/30\relax}
\ifdim\tabcolsep<1.5pt\setlength{\tabcolsep}{1.5pt}\fi
\typeout{EBTAB tab_rq5a natural=\the\wd0\space line=\the\linewidth\space sep=\the\tabcolsep}
\ebtab
\end{table*}

\smallskip\noindent\textbf{Completion with modeled execution.}
Decision latency decides how much load an interpreter can admit
(Fig.~\ref{fig:rq5a}a and Table~\ref{tab:rq5a}). With four interpretation
slots, \jev completes 0.907--0.953 of the requests exactly correct and on time
at every offered load from 1 to 16 requests/s. \deepseek holds 0.907--0.983 up
to 4 requests/s and falls to 0.603 at 8 and 0.093 at 16. \glm falls below 0.1
from 8 requests/s on, and \qwenflash from 4. The completion gap to \jev
therefore grows with load for all three LLMs. Per doubling of $\lambda$ the
gap widens by 0.19, 0.25, and 0.24. The 95\% confidence intervals are
$[0.12, 0.25]$, $[0.22, 0.27]$, and $[0.22, 0.25]$, with Holm-adjusted
$p<0.001$ (H6). At low load, \deepseek completes slightly
more requests exactly than \jev (0.983 against 0.937 at 4 requests/s), because
\jev's remaining semantic errors leave execution unchanged. Counting operational
completion, \jev reaches 0.97--1.00.

Deadlines tell the same story from the other side (Fig.~\ref{fig:rq5a}c). At a
0.5\,s deadline, \jev still completes 0.890 and \deepseek 0.377, while \glm and
\qwenflash complete no request. At 4\,s, \deepseek and \glm reach 0.987 and
0.990. Bursty arrivals at the center rate separate the LLMs further. \deepseek
keeps 0.987 and \jev 0.950, while \glm drops from 0.807 to 0.223 and
\qwenflash stays near zero at 0.063. On the edge node,
\semif and \qwenjson saturate from 2 requests/s on, since one accelerator
serves the four slots in turn. Their gap therefore narrows with load (slope
$-0.09$), and the self-hosted analogue of H6 does not hold. \laya stays fast
but completes only 0.093--0.117 exactly, limited by its accuracy. Varying the number
of edge nodes from 5 to 40 changes completion by less than one point when the
same decision timeline is replayed. Hence the swings between the live topology
cells reflect provider latency at the time of each run.

\begin{figure*}[!t]
\centering
\includegraphics[width=7.00in]{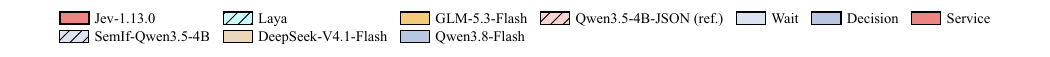}\\[-2pt]
\subfloat[Correct completion, cache off]{\includegraphics[width=1.70in]{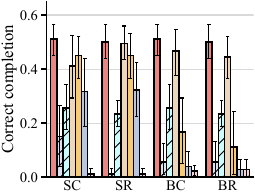}\label{fig:rq5b_a}}\hfill
\subfloat[Correct completion, cache on]{\includegraphics[width=1.70in]{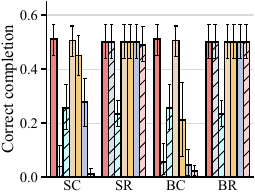}\label{fig:rq5b_b}}\hfill
\subfloat[Latency p95, cache off]{\includegraphics[width=1.70in]{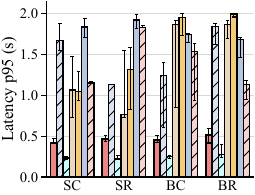}\label{fig:rq5b_c}}\hfill
\subfloat[Time breakdown (SC0)]{\includegraphics[width=1.70in]{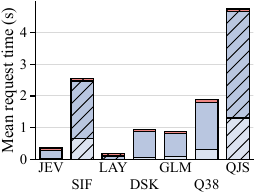}\label{fig:rq5b_d}}
\caption{RQ5, real OCR service (Part~B). (a) Correct-completion rate with cache off (whiskers are 95\% confidence intervals) and (b) cache on across operational conditions: S steady, B bursty, C changing text, R repeated text. (c) 95th percentile request latency $T$ on completions with cache off. (d) Stacked mean time decomposition into admission wait, decision, and service execution for SC0. Model codes: JEV \jev, SIF \semif, LAY \laya, DSK \deepseek, GLM \glm, Q38 \qwenflash, QJS \qwenjson{} (ref.).}
\label{fig:rq5b}
\end{figure*}

\begin{table*}[!t]
\centering
\caption{RQ5, real OCR service (Part~B). Correct completion rate and 95th percentile request latency $T$ on completions across operational conditions under steady and bursty arrivals. Model codes: JEV \jev, SIF \semif, LAY \laya, DSK \deepseek, GLM \glm, Q38 \qwenflash, QJS \qwenjson{} (ref.). Best value per row within each metric block in bold (QJS reference excluded). 95\% confidence intervals are drawn in Fig.~\ref{fig:rq5b}.}
\label{tab:rq5b}
\scriptsize
\setlength{\tabcolsep}{3pt}
\setlength{\aboverulesep}{0pt}\setlength{\belowrulesep}{0pt}\setlength{\extrarowheight}{0pt}
\renewcommand{\arraystretch}{0.94}
\def\ebtab{%
\begin{tabular}{lrrrrrrrrrrrrrr}
\toprule
\rowcolor{ebBlue} & \multicolumn{7}{c}{Correct completion $\uparrow$} & \multicolumn{7}{c}{p95 $T$ (s) $\downarrow$} \\
\cmidrule(lr){2-8}\cmidrule(lr){9-15}
\rowcolor{ebBlue}Condition & JEV & SIF & LAY & DSK & GLM & Q38 & QJS & JEV & SIF & LAY & DSK & GLM & Q38 & QJS \\
\midrule
\rowcolor{ebAmber!45}\multicolumn{15}{l}{\textbf{\textit{Steady arrivals}}} \\
changing, off & \cellcolor{ebCoral!35}\textbf{0.511} & 0.150 & 0.256 & 0.411 & 0.450 & 0.317 & 0.011 & \cellcolor{ebCoral!10}0.42 & 1.68 & \cellcolor{ebCoral!35}\textbf{0.23} & 1.07 & 1.05 & 1.84 & 1.15 \\
changing, on & \cellcolor{ebCoral!35}\textbf{0.511} & 0.039 & 0.256 & 0.506 & 0.450 & 0.278 & 0.011 & \cellcolor{ebCoral!10}0.48 & 1.72 & \cellcolor{ebCoral!35}\textbf{0.22} & 0.69 & 1.57 & 1.95 & 1.20 \\
repeated, off & \cellcolor{ebCoral!35}\textbf{0.500} & 0.011 & 0.233 & 0.494 & 0.450 & 0.322 & 0.011 & \cellcolor{ebCoral!10}0.47 & 1.14 & \cellcolor{ebCoral!35}\textbf{0.22} & 0.76 & 1.32 & 1.92 & 1.83 \\
repeated, on & \cellcolor{ebCoral!35}\textbf{0.500} & \cellcolor{ebCoral!35}\textbf{0.500} & 0.233 & \cellcolor{ebCoral!35}\textbf{0.500} & \cellcolor{ebCoral!35}\textbf{0.500} & \cellcolor{ebCoral!35}\textbf{0.500} & 0.489 & \cellcolor{ebCoral!10}0.13 & 0.13 & 0.13 & \cellcolor{ebCoral!35}\textbf{0.12} & 0.13 & 0.13 & 0.13 \\
\rowcolor{ebAmber!45}\multicolumn{15}{l}{\textbf{\textit{Bursty arrivals}}} \\
changing, off & \cellcolor{ebCoral!35}\textbf{0.511} & 0.056 & 0.256 & 0.467 & 0.167 & 0.039 & 0.022 & \cellcolor{ebCoral!10}0.46 & 1.24 & \cellcolor{ebCoral!35}\textbf{0.25} & 1.86 & 1.95 & 1.75 & 1.54 \\
changing, on & \cellcolor{ebCoral!35}\textbf{0.511} & 0.056 & 0.256 & 0.506 & 0.211 & 0.044 & 0.022 & \cellcolor{ebCoral!10}0.45 & 1.26 & \cellcolor{ebCoral!35}\textbf{0.26} & 1.36 & 1.96 & 1.98 & 1.54 \\
repeated, off & \cellcolor{ebCoral!35}\textbf{0.500} & 0.056 & 0.233 & 0.444 & 0.111 & 0.028 & 0.028 & \cellcolor{ebCoral!10}0.52 & 1.84 & \cellcolor{ebCoral!35}\textbf{0.28} & 1.87 & 2.00 & 1.69 & 1.13 \\
repeated, on & \cellcolor{ebCoral!35}\textbf{0.500} & \cellcolor{ebCoral!35}\textbf{0.500} & 0.233 & \cellcolor{ebCoral!35}\textbf{0.500} & \cellcolor{ebCoral!35}\textbf{0.500} & \cellcolor{ebCoral!35}\textbf{0.500} & 0.500 & \cellcolor{ebCoral!10}0.14 & 0.16 & \cellcolor{ebCoral!35}\textbf{0.10} & 0.14 & 0.15 & 0.13 & 0.15 \\
\bottomrule
\end{tabular}}
\def\ebh#1#2{#1 #2}\sbox0{\ebtab}
\ifdim\wd0>\dimexpr\linewidth-1pt\relax\def\ebh#1#2{\shortstack{#1\\{}#2}}\sbox0{\ebtab}\fi
\setlength{\tabcolsep}{\dimexpr\tabcolsep+(\linewidth-\wd0-1pt)/30\relax}
\ifdim\tabcolsep<1.5pt\setlength{\tabcolsep}{1.5pt}\fi
\typeout{EBTAB tab_rq5b natural=\the\wd0\space line=\the\linewidth\space sep=\the\tabcolsep}
\ebtab
\end{table*}

\smallskip\noindent\textbf{Correct completion in the real OCR service.}
With real recognition, the recognizer sets the ceiling
(Fig.~\ref{fig:rq5b}a--b and Table~\ref{tab:rq5b}). \jev dispatches all 180
OCR requests of every condition to a permitted node and tier, and each
finishes before its deadline. Tesseract reads 92 of the 180 changing-text images exactly and 90 of
the repeated ones. \jev's correct completion of 0.511 and 0.500 therefore
equals the recognizer's own accuracy. Under steady arrivals with changing text, \deepseek
completes 0.411, \glm 0.450, and \qwenflash 0.317. Bursty arrivals cut \glm to
0.167 and \qwenflash to 0.039, while \deepseek keeps 0.467 and \jev 0.511. On
the edge node, \semif completes 0.150 and \qwenjson 0.011, both held back by
the shared accelerator, and \laya completes 0.256. Because the two changing-text
conditions differ only in a cache that never hits, they act as a repeated
run. \jev reproduces 0.511, and \deepseek moves from 0.411 to 0.506 with
provider latency between the runs.

No interpreter dispatched an unsupported request to OCR or sent an image to a
node its locality field forbids. \jev rejected all 60 counting and detection
requests in every condition. The slower interpreters lost part of these
rejections to the same queue expiry that costs them OCR completions. Under
bursty changing text, \glm rejected 20 of the 60 in time and \qwenflash 6.

\smallskip\noindent\textbf{End-to-end latency and the cache boundary.}
On shared successes, the requests that both \jev and an LLM complete
correctly, \jev's 95th-percentile request latency is lower in every eligible
cell of both parts (H7). The difference is 0.06--1.41\,s against \deepseek in
16 of 16 cells and 0.50--1.52\,s against \glm in 11 of 11. Because \qwenflash shares
at least 30 successes with \jev in only four cells, one short of the five that
H7 requires, H7 is not evaluable for it. In those four cells its p95 is
1.29--1.48\,s higher. The gap remains after image transfer, worker queueing, and
recognition. Under steady changing text in Part~B, the p95 differences are
0.64\,s for \deepseek, 0.63\,s for \glm, and 1.42\,s for \qwenflash
(Fig.~\ref{fig:rq5b}c). Recognition itself takes 0.07--0.08\,s on average,
so decision time and the admission wait it causes make up most of each
request (Fig.~\ref{fig:rq5b}d).

Caching repeated descriptions removes most of this difference (H8). With
eight recurring descriptions, the cache answers 267--292 of 300 Part~A
requests and 229--232 of 240 Part~B requests, and the median request latency
of every interpreter except \laya falls to 0.06--0.09\,s. The median-latency
gap to \jev shrinks in all nine contrasts, by 0.11--1.31\,s. In both parts the completion gap closes wherever one existed.
The gap shrinks by 0.22 for \glm and 0.94 for \qwenflash in the Part~A reuse
block (Fig.~\ref{fig:rq5a}d), and in Part~B every interpreter except
\laya reaches 0.489--0.500 with the cache on. The three completion contrasts
that do not resolve start from gaps of at most 0.05 without the cache, where
\deepseek and \glm already match \jev. H8 therefore holds in full for \glm and
\qwenflash in Part~A and for \qwenflash in Part~B, and holds for median
latency for all three LLMs. \laya stays at 0.233 with or without the cache,
which replays its decisions unchanged.

\smallskip\noindent\textbf{Billed API cost.}
Fees per correct completion follow completion, because every billed call
counts, including those whose request later fails. At 1 request/s, \jev's API
fees are 0.037\,USD per 1{,}000 correct completions, against 0.115 for
\deepseek, 0.043 for \glm, and 0.058 for \qwenflash. At 16 requests/s \jev
stays at 0.038\,USD, while the three LLMs reach 0.58, 0.65, and 2.13\,USD. In
Part~B under steady changing text, the fees are 0.090\,USD for \jev against
0.309, 0.117, and 0.217\,USD. These fees are higher than in Part~A because every interpreter
also pays for decisions whose recognition fails. With repeated descriptions
cached, all four fall to 0.001--0.009\,USD. Of the 19{,}410 hosted calls, 24
reported no usage and are left out of the fees.

\takeaway{End-to-end service}{Decision time decides how much load an
admission path can take. \jev keeps 0.91--0.95 exact, on-time completion from 1
to 16 requests/s, where the LLMs fall below 0.1. In the real OCR service
it reaches the recognizer's own accuracy under every arrival pattern. Because caching
repeated descriptions gives the interpreters nearly the same latency, \jev's
advantage lies in requests that need a fresh decision.}

\section{Discussion}
\label{sec:discussion}
\subsection{Implications for Service Orchestration}
The experiments show a practical use for Jev in edge-service admission,
where it reduces the time and API fees spent interpreting requests on bounded
contracts. The shared intent contract lets
Jev replace the generative interpreter within the same validation and
scheduling pipeline. The real OCR results show that the latency
benefit remains visible after image transfer, worker queueing, and
recognition. On requests that both complete, \jev's 95th-percentile request
latency stays 0.63--1.42\,s below the LLMs' under steady arrivals with
changing text.

The interpretation layer also marks where the substitution stops paying.
\jev's decision time barely moves with input length, bundling, contract
size, or catalog size. Hence its advantage grows where generative
interpreters slow down. Its accuracy does move with contract size. With four
fields it trails the strongest LLM by a few points, and with eight by more
than 30. A service designer can therefore place \jev on short contracts and
latency-bound paths, and route wide contracts to a generative interpreter.
Request-time catalogs favor both families over a trained classifier, since
neither needs labelled examples for a new service.

This integration gives service designers two ways to reduce interpretation
overhead. A faster decision backend shortens fresh calls, while caching
reuses decisions for repeated descriptions. With eight recurring descriptions,
the cache answers nearly every request. Every interpreter except \laya,
including the self-hosted \qwenjson, then completes 0.489--0.500, the
recognizer's accuracy.
Choosing between backends therefore depends on the share of requests that
need a new decision and the time available for service execution. In the
admission path studied here, shorter decision waiting leaves more deadline
slack and releases interpretation capacity sooner.

\subsection{Evaluating the Complete Service Path}
The shared scheduler connects interpreter performance to delivered service.
It reads current worker state after interpretation and checks placement,
tier, priority, and predicted completion time before dispatch. Evaluating
intent fidelity, execution compliance, and actual output separately makes
it possible to trace how an interpretation affects the final result.
For example, the normal-versus-unspecified urgency difference maps to the
same execution priority, while the returned OCR text determines whether
recognition succeeded. This evaluation supports a service-level choice of
interpreter using completion, response time, and fees together.

\subsection{Scope and Next Steps}
Because \corpus is generated and verified by language models, its wording may
favor phrasing that such models parse easily. Programmatic padding and noise
conditions and the blind agreement filter limit this effect. Each case
is run once per interpreter, and the service layer uses one arrival trace per
cell. The real service has three workers and one service
family. Hosted timings include provider and network paths, with sequential
runs exposed to temporal variation. They compare deployed services rather
than isolate internal model inference. API fees cover interpretation calls,
whereas local compute and communication costs are outside this measure.
Section~\ref{sec:settings} records the configuration.

The next step is to evaluate independently authored requests across days,
services, and network conditions, using deadlines drawn from application
requirements. Optimized local serving, a bounded extractor, and decision models with
larger option limits than \semif and \laya would extend the available
deployment choices.

\section{Conclusion}
\label{sec:conclusion}
We integrated \jev into an edge-service admission pipeline and compared it with
two self-hosted decision models and three hosted LLMs on \corpus and on a live
admission path. \jev's median decision latency is 22.7--64.5\% below that of
\deepseek and barely moves with input length, bundling, contract size, or
catalog size. On four-field contracts, \jev also cuts API fees per correct
decision by 59.7--80.9\% and trails \deepseek by a few exact-match points. Wide
contracts bound the substitution, because the accuracy gap exceeds 30 points at
eight fields. Request-time catalogs let \jev name unseen services as accurately
as known ones. On the live admission path, \jev keeps 0.91--0.95 of requests
exact and on time at up to 16 requests/s, whereas the LLMs fall below 0.1.
Caching repeated descriptions gives the interpreters nearly the same latency,
which confines \jev's advantage to requests that need fresh interpretation. A
service designer can therefore use \jev for latency-bound admission on short
contracts, pair it with caching for recurring requests, and route wide
contracts to a generative interpreter.


\end{document}